\documentstyle[preprint,revtex]{aps}

\newcommand{\half}{{\case{1}/{2}}}

\newcommand{\onasq}{{\case{1}/{a^{2}}}}
\newcommand{\xonyasq}[2]{{\case{{#1}}/{{#2}a^{2}}}}
\newcommand{\ids}{\int d\sigma \:}
\newcommand{\idsx}[1]{\int d\sigma{#1} \:}
\newcommand{\idssp}{\ids d\sigma' \:}
\newcommand{\onal}{{\case{1}/{\alpha}}}
\newcommand{\hlsq}{\half L^{2}}
\newcommand{\vol}{a^{n} \Omega_{n+1}}

\newcommand{\schm}[3]{Y_{l{#1}m{#2}}(\eta{#3})}
\newcommand{\cschm}[3]{Y^{\ast}_{l{#1}m{#2}}(\eta{#3})}
\newcommand{\dlz}{(2l+n-1)\;\; \frac{\Gamma(l+n-1)}{\Gamma(n) \Gamma(l+1)}}
\newcommand{\sphm}[5]{Y^{({#4})}_{l{#1}s{#2}{#5}}(\eta{#3})}
\newcommand{\csphm}[5]{Y^{\dagger \,({#4})}_{l{#1}s{#2}{#5}}(\eta{#3})}
\newcommand{\sigdl}{\Sigma \cdot L}
\newcommand{\sigdtl}[1]{\left( \Sigma \cdot T_{l{#1}} \right)}
\newcommand{\gamdeta}[1]{\gamma \cdot \eta{#1}}
\newcommand{\gamdD}[1]{\gamma \cdot D{#1}}
\newcommand{\dlone}{(n-1)\frac{l(l+n-1)}{(l+1)(l+n-2)}\;\; \dlz }
\newcommand{\vhm}[6]{Y^{({#4}){#5}}_{l{#1}m{#2}{#6}}(\eta{#3})}
\newcommand{\tvhm}[5]{Y^{(0){#4}}_{l{#1}M{#2}{#5}}(\eta{#3})}
\newcommand{\cvhm}[6]{Y^{\ast ({#4}){#5}}_{l{#1}m{#2}{#6}}(\eta{#3})}

\newcommand{\tlm}[5]{\left( T_{l{#1}} \, ^{{#2}}_{{#3}} \right)_{m{#4}\,
m{#5}}}
\newcommand{\vlm}[5]{\left( V_{l{#1}} \, ^{{#2}}_{{#3}} \right)_{m{#4}\,
m{#5}}}
\newcommand{\ulm}[5]{\left( U_{l{#1}} \, ^{{#2}}_{{#3}} \right)_{m{#4}\,
m{#5}}}
\newcommand{\wlm}[5]{\left( W_{l{#1}} \, ^{{#2}}_{{#3}} \right)_{m{#4}\,
M{#5}}}
\newcommand{\cwlm}[5]{\left( W_{l{#1}} \, ^{{#2}}_{{#3}}
\right)^{\ast}_{M{#4}\, m{#5}}}
\newcommand{\chilms}[5]{\left( \chi_{l{#1}}^{({#2})} \, _{{#3}}
\right)_{m{#4}\, s{#5}}}
\newcommand{\tl}[3]{\left( T_{l{#1}} \, ^{{#2}}_{{#3}} \right)}
\newcommand{\vl}[3]{\left( V_{l{#1}} \, ^{{#2}}_{{#3}} \right)}
\newcommand{\ul}[3]{\left( U_{l{#1}} \, ^{{#2}}_{{#3}} \right)}
\newcommand{\wl}[3]{\left( W_{l{#1}} \, ^{{#2}}_{{#3}} \right)}
\newcommand{\chil}[2]{\left( \chi_{l{#1}}^{({#2})} \right)}
\newcommand{\etal}[4]{\left( \eta_{l{#1}}^{({#2})} \, ^{{#3}}_{{#4}} \right)}
\newcommand{\Dl}[4]{\left( D_{l{#1}}^{({#2})} \, ^{{#3}}_{{#4}} \right)}
\newcommand{\Ql}[4]{\left( Q_{l{#1}}^{({#2})} \, ^{{#3}}_{{#4}} \right)}
\newcommand{\htlsq}[1]{\left( \half T_{l{#1}}^{2} \right)}

\newcommand{\p}[3]{ P_{l{#1}}^{({#2})}(\eta{#3})}
\newcommand{\vp}[5]{ P_{l{#1}}^{(#2)} \, ^{{#3}}_{{#4}} (\eta{#5})}
\newcommand{\vpl}[4]{ \left( P_{l{#1}}^{(#2)} \, ^{{#3}}_{{#4}} \right)}
\newcommand{\Q}[3]{ Q^{{#1}}_{{#2}}(\eta {#3})}
\newcommand{\Qop}[2]{ Q^{{#1}}_{{#2}}}
\newcommand{\lop}[2]{ L^{{#1}}_{{#2}}}
\newcommand{\vpop}[4]{ P_{l{#1}}^{(#2)} \, ^{{#3}}_{{#4}}}
\newcommand{\pop}[2]{ P^{{#1}}_{{#2}}}
\newcommand{\kdelta}[2] { \delta^{{#1}}_{{#2}}}
\newcommand{\slsh}[1]{{\not \! #1}}

\newcommand{\nontwo}{{\case{n}/{2}}}
\newcommand{\eps}{\varepsilon}
\newcommand{\Aterm}{\nontwo - 1}
\newcommand{\oneps}{{\case{1}/{\eps}}}
\newcommand{\twooneps}{{\case{2}/{\eps}}}
\newcommand{\FactorOne}{{\case{e^{2} (\mu)^{4-n}}/{n \vol}}}
\newcommand{\FactorTwo}{{\case{e^{2} (\mu a)^{4-n}}/{n \Omega_{n+1} a^{2}}}}
\newcommand{\FactorThree}{{\case{ 2^{n/2} e^{2} (\mu a)^{4-n}}/{n \Omega_{n+1}
a^{2}}}}
\newcommand{\FactorFour}{{\case{e^{2}}/{16 \pi^{2} a^{2}}}}
\newcommand{\FactorFive}{{\case{e^{2} (\mu a)^{4-n}}/{n(n-1) \Omega_{n+1}
a^{2}}}}
\newcommand{\FactorSix}{{\case{e^{2}}/{12 \pi^{2} a^{2}}} \left( \twooneps
\right)}
\newcommand{\FactorSixA}{{\case{e^{2}}/{12 \pi^{2}}}  \left( \twooneps \right)}
\newcommand{\FactorSeven}{{\case{e^{2} (\mu a)^{4-n}}/{2^{n/2} \Omega_{n+1}}}}
\newcommand{\FactorSevenA}{{\case{1}/{2^{n/2} \vol }}}
\newcommand{\FactorEight}{{\case{e^{2} (\mu a)^{4-n}}/{\Omega_{n+1}}}}
\newcommand{\FactorNine}{{\case{e^{2}}/{16\pi^{2}}} \alpha \left( \twooneps
\right)}
\newcommand{\FactorTen}{{\case{e^{2} (\mu a)^{4-n}}/{2^{n/2} n \Omega_{n+1}}}}
\newcommand{\FactorEleven}{{\case{e^{2} (\mu a)^{4-n}}/{ n \Omega_{n+1}}}}
\newcommand{\spop}[1]{ \left[ (\Aterm) - \sigdl{#1} \right]}
\newcommand{\spopl}[1]{ \left[ (\Aterm) - \sigdtl{{#1}} \right]}
\newcommand{\spoplab}[3]{\left[ (\Aterm)\kdelta{{#2}}{{#3}}-
\tl{{#1}}{{#2}}{{#3}} \right] }
\newcommand{\vctop}[4]{ \left[ \alpha \kdelta{#2}{#3} + (1-\alpha)
\vp{#1}{0}{#2}{#3}{#4} \right]}
\newcommand{\vctopl}[3]{ \left[ \alpha \kdelta{#2}{#3} + (1-\alpha)
\vpl{#1}{0}{#2}{#3} \right] }
\newcommand{\idsthree}[3]{ \int d\sigma{#1} d\sigma{#2} d\sigma{#3}}

\newcommand{\inda}{\hspace{1em}}
\newcommand{\indb}{\hspace{2em}}
\newcommand{\indc}{\hspace{5em}}
\newcommand{\indd}{\hspace{7em}}
\newcommand{\inde}{\hspace{10em}}

\begin{document}
\draft
\begin{title}
A Self-Consistent Formulation of
Quantum Field Theory on $S_{4}$.
\end{title}
\author{B.A.Harris and G.C.Joshi}
\begin{instit}
School of Physics,
University Of Melbourne,
Parkville, Victoria 3052, Australia
\end{instit}
\begin{abstract}
Recent developments in quantum gravity suggest that wormholes may influence the
observed values
of the constants of nature. The Euclidean formulation of quantum gravity
predicts that wormholes
induce a probability distribution in the space of possible fundamental
constants.
This distribution may computed by evaluating the functional
integral about the stationary points of the action.
In particular, the effective action on a large spherical
space may lead to the vanishing of the cosmological constant and possibly
determine the
values of other constants of nature. The ability to perform calculations
involving interacting
quantum fields, particularly non-Abelian models, on a four-sphere is vital if
one
is to investigate this possibility.
In this paper we present a self-consistent formulation
of field theory on a four-sphere using the angular momentum space
representation of $SO(5)$.
We give a review of field theory on a sphere and then show how a matrix element
prescription
in angular momentum space overcomes previous limitations in calculational
techniques.  The standard
one-loop graphs of QED are given as examples.
\end{abstract}
\newpage
\section{Introduction}

Recently it has been suggested\cite{col1} by Coleman that the effect of
wormholes in euclidean quantum gravity is to modify or even determine the
observed values of the fundamental constants.  In this idea, the effects of
wormholes can be described by as set of parameters $\{\alpha_i\}$ in which the
values
of constants such as the cosmololgical constant $\Lambda$, the gravitational
constant $G$, and other couplings depend at the scale of wormhole physics, at
 or just below the Planck scale.  Furthermore, Coleman has suggested a
mechanism by which $\Lambda$ is exactly zero and $G$ assumes its minimum
possible
value.  In the euclidean functional integral, it is possible\cite{col2}
integrate out the contribution of wormhole configurations to give an integral
over a
distribution of a parameters.  It is suggested that the dominant contribution
comes from classical saddle points of the integral.  The distribution is of the
form
\begin{equation}
	Q(\alpha) = e^{e^{-\Gamma_{eff}\{\alpha_i\}}}
\end{equation}
For the action
\begin{equation}
	\Gamma[g] = \int d^{4}x \: \left[\Lambda - {\case{R}/{16\pi G}} +
		aR_{\mu \nu \sigma \lambda}R^{\mu \nu \sigma \lambda} +
		bR_{\mu \nu}R^{\mu \nu} + cR^{2} \right] ,
\end{equation}
 with $\Lambda>0$ , the stationary point is a four-sphere of radius
\begin{equation}
	a = \sqrt{{\case{3}/{8G\Lambda}}}\; ,
\end{equation}
with
\begin{equation}
	\Gamma_{eff} = - {\case{3}/{8G^{2}\Lambda}} + {\case{8\pi^{2}d}/{3}}\; ,
	\label{effact}
\end{equation}
where $d$ is a linear combination of $a$,$b$ and $c$.
The double exponential makes the $\Lambda=0$ surface in $\alpha$-space
overwhelmingly likely and the minimization of $G$ and
$d$ on that surface should fix some or all of the other constants of nature.
Since the higher derivative terms  are related to the trace
anomaly\cite{shore4}, there
have been attempts\cite{grin1,grin2} to use renormalization group techniques to
relate the
coupling constants of interacting field theories to the constant $d$.  A
consistent method to calculate amplitudes on $S_{4}$ is required,
especially if the program is to be taken beyond one loop to investigate the
effect of vacuum
diagrams containing interactions.  This paper contains a brief review of the
formalism of field theory on $S_{4}$, and then describes a self consistent way
to
extract amplitudes of interest using angular momentum representations of
$SO(5)$. The method presented here to compute one-loop
Feynman diagrams, we believe, is a major improvement allowing
amplitudes of greater complexity, such as those encountered in non-Abelian
models, to be handled.
While the previous results will be reproduced, the method is straightforward
and removes the need
to come up with specific tricks to handle individual calculations.

\section{Field Theory in Spherical Spacetime}

This review closely follows work published in a series of papers by
Drummond and Shore\cite{drum1,ds1,ds2,ds3} with slight changes of notation
where convienient.
The form of the action may be deduced by
conformally mapping the flat euclidean four-space action to the surface of a
hypersphere embedded in five dimensions.  To do this it is necessary to
introduce a fifth co-ordinate $x_{5}$ and a unit vector $h$ such that the $x
\cdot h = 0$
plane is identified with the original space.  The conformal transformation
\begin{equation}
	g_{ab} = \kdelta{}{ab} \rightarrow \kappa^{2} \kdelta{}{ab} ,
\end{equation}
where
\begin{equation}
	\kappa = 2/ \left(1 - {\case{2 x \cdot h}/{a}} + \xonyasq{x^{2}}{}\right) ,
\end{equation}
maps the plane $x \cdot h = 0$ onto the sphere $r = a$.  One now performs a
general
co-ordinate transformation to the flat five dimensional co-ordinates $\eta$
given by
\begin{equation}
	\eta^{a} = ah^{a} + \kappa \left(x^{a} - {\case{x^{2}}/{a}} h^{a}\right) .
\end{equation}
For scalar fields this process is straightforward, but for spin $1/2$
and spin $1$ fields it is more involved.

For fermions it is necessary to define a vierbein and a spin connection.
These objects transform in the usual way under conformal and general
co-ordinate transformations.  The additional matrix $\gamma_{5}$ is needed but
it can be
shown that since $\gamma$ matrices always contract with a spherical projection
operator $Q$, there are effectively only four independent 'gamma' matrices.
In addition, the combination $\gamma \cdot \eta /a$ effectively acts as
$\gamma_{5}$, anticommuting with
all of the $Q$-contracted gamma matrices.   This justifies the mapping of the
four component $SO(5)$ spinors to ordinary Dirac spinors on the surface,
differing from the approach of Adler\cite{adler}.

For a vector particle, a similar procedure with an additional $A_{5}$ field
allows one to derive the spherical action and show that only four
independent components ( not counting gauge and  dynamical constraints )
survive.

Since dimensional regularization will be used throughout, the sphere
will be $n$ dimensional and the space indices will run from $1$
to $(n+1)$ with $n = 4 - \eps$.

At this point it is useful to define a some of the operators used on the
sphere.  The rotation generators
\begin{equation}
	\lop{}{ab} = \eta_{a} \partial_{b} - \eta_{b} \partial_{a} ,
\end{equation}
and
\begin{equation}
	\hlsq  = \half \lop{}{ab} \lop{ab}{} ,
\end{equation}
define invariant derivatives on the sphere.  The tangential projection
operator
\begin{equation}
	\Q{}{ab}{} = \kdelta{}{ab} - \onasq \eta_{a} \eta_{b} ,
	\label{Qdef}
\end{equation}
enables one to define a tangential derivative
\begin{equation}
	D_{a} = \Q{}{ab}{} \partial^{b} = \onasq \eta_{b} \lop{b}{\;\; a},
\end{equation}
which satisfies the identities
\begin{eqnarray}
	\eta_{a} D^{a} & = & 0 ,\nonumber \\
	D^{a} \eta_{a} & = & n ,
\end{eqnarray}
and
\begin{equation}
	D^{2} = \xonyasq{1}{2} L^{2} .
\end{equation}

\section{Classical Actions}

The simplest example is the self interacting scalar field.  The
action\cite{drum1}
for this case is
\begin{equation}
	S_{E} = \ids \left[ - \half \phi M \phi + V(\phi) \right] ,
		\label{Sact}
\end{equation}
where the kinetic operator
\begin{equation}
	M = D^{2} - \xonyasq{1}{4} n(n-2) ,
\end{equation}
shows the correspondence between $\partial$ and $D$ and also exhibits the
conformal
term $-\phi^{2}R/6$.

For a free massless Dirac fermion the action is\cite{ds1}
\begin{equation}
	S_{E} =  \ids \left[ \bar{\psi} \gamma^{a}
		\left( D_{a} - \xonyasq{n}{2} \eta_{a} \right) \psi \right] ,
\end{equation}
where the $\gamma$ matrices satisfying
\begin{equation}
	\left\{ \gamma^{a} , \gamma^{b} \right\} = 2 \kdelta{ab}{} ,
\end{equation}
\begin{equation}
	\left\{ \Q{ab}{}{} \gamma_{b} , \Q{cd}{}{} \gamma_{d} \right\}  = 2 \Q{ac}{}{}
,
\end{equation}
and
\begin{equation}
	\left\{ \Q{ab}{}{} \gamma_{b} \, , \,  \gamdeta{} \right\} = 0.
\end{equation}
Note that the term involving $\gamdeta{}$ is necessary
to ensure invariance of the action under
\begin{eqnarray}
	\psi & \rightarrow & e^{-i\theta(\gamdeta{} / a)} \psi, \\
	\bar{\psi} & \rightarrow & \bar{\psi} e^{-i\theta(\gamdeta{} /a)},
\end{eqnarray}
which is the spherical version of a chiral transformation.

The action for a massless gauge boson is\cite{ds1}
\begin{equation}
	S_{E} = \ids \left[ \xonyasq{1}{12} F_{abc} F^{abc} \right] ,
\end{equation}
where
\begin{equation}
	F_{abc} = \lop{}{ab}A_{c} + \lop{}{bc}A_{a} + \lop{}{ca}A_{b}.
\end{equation}
Integrating by parts gives
\begin{equation}
	S_{E} = \ids \left[ - \xonyasq{1}{2} A^{a} \pop{}{ab} A^{b} \right] ,
		\label{Vact}
\end{equation}
with $\pop{}{ab}$ given by
\begin{equation}
	\pop{}{ab} = \hlsq  \kdelta{}{ab} + \lop{\;\; c}{a} \lop{}{cb} -
(n-1)\lop{}{ab} .
\end{equation}
The spin $1$ projection operator $\pop{}{ab}$ satisfies the following
identities
\begin{equation}
	D_{a}\pop{ab}{} = \pop{ab}{}D_{b} = \eta_{a}\pop{ab}{} = \pop{ab}{}\eta_{b} =
0, \label{P1}
\end{equation}
\begin{equation}
	\pop{}{ab}\lop{b}{\;\; c} = \lop{\;\; b}{a} \pop{}{bc} = \pop{}{ac} ,
\label{P2}
\end{equation}
\begin{equation}
	\pop{}{ab}\pop{b}{\;\; c} = N \pop{}{ac} = \pop{}{ac} N , \label{P3}
\end{equation}
where
\begin{equation}
	N = \hlsq  - n + 2.
\end{equation}
It turns out that Eq.(\ref{Vact}) is the more useful version of the action when
the
theory is quantised.

Gauge invariant actions may be constructed through the substitutions
\begin{equation}
	D^{a} \rightarrow D^{a} - ie \Q{ab}{}{} A_{b}
\end{equation}
or
\begin{equation}
	\lop{}{ab} \rightarrow \lop{}{ab} - ie( \eta_{a} A_{b} - \eta_{b} A_{a}).
\end{equation}
This allows one to write down the action for scalar and spinor
electrodynamics
\begin{eqnarray}
	S_{E} & = & S_{E}^{free} + \ids \left[ -ie \, \bar{\psi} \Q{ab}{}{}
			\gamma_{b} \psi A_{a} \right. \nonumber \\
	& & \inda \mbox{} + \onasq ie \, \eta_{a} A_{b} ( \phi^{*}\lop{ab}{} \phi
			 - \lop{ab}{} \phi^{*} . \phi) \nonumber \\
	& & \inda \mbox{} + \left. e^{2} \, \Q{}{ab}{} A^{a} A^{b} \phi^{*} \phi
\right] ,
\end{eqnarray}
which is invariant under the local transformation
\begin{eqnarray}
	\psi & \rightarrow & e^{ie\Lambda(\eta)} \psi , \nonumber \\
	\phi & \rightarrow & e^{ie\Lambda(\eta)} \phi , \nonumber \\
	A^{a} & \rightarrow & A^{a} + D^{a}\Lambda(\eta) . \label{gauge}
\end{eqnarray}
There is an additional invariance of the vector particle action under
\begin{equation}
	A^{a} \rightarrow A^{a} + \onasq \eta^{a} \theta(\eta), \label{radial}
\end{equation}
which reflects the fact that only tangential components of the vector field
contribute to the action.  The above procedure may be generalized to non-
Abelian symmetries in the usual way. Finally,  Yukawa
and scalar self couplings have the same form as in flat space.

\section{Quantization}

\label{Quantization}

Quantization of the theory is achieved through the functional integral
method.  The simplest example is the interacting scalar field.  Define the
generating functional\cite{drum1}
\begin{equation}
	Z[J] = \int {\cal D} \phi \; \exp \left\{ - S_{E} + \ids J(\eta) \phi(\eta)
\right\} ,
\end{equation}
where $S_{E}$ is given by Eq.(\ref{Sact}). This can be rewritten as
\begin{equation}
	Z[J] = \exp \left\{ - \ids V\left(\frac{\delta}{\delta J(\eta)} \right)
\right\}
		Z_{0}[J] ,
\end{equation}
where
\begin{equation}
	Z_{0}[J] = \int {\cal D} \phi \; \exp \left\{ \ids
		\left( \half \phi M \phi + J \phi \right) \right\} ,
\end{equation}
Completing the square and performing the Gaussian integral gives
\begin{equation}
	Z_{0}[J] = const.\, [det(-M)]^{-\half} exp \left\{ -\half \ids
		d\sigma' \: J(\eta) M^{-1}(\eta,\eta') J(\eta') \right\}. \label{Sfgen}
\end{equation}
The propagator is defined by functional differentiation,
\begin{equation}
	G(\eta,\eta') = \left. \frac{\delta^{2} Z_{0}[J]}{\delta J(\eta) \delta
J(\eta')} \right|_{J=0}
		= - M^{-1}(\eta,\eta') ,
		\label{scprop}
\end{equation}
where $M^{-1}$ may be expressed in either coordinate space form\cite{drum1} or
in the more
useful angular momentum space representation.

Given a particular form of potential,  for example,
\begin{equation}
	V(\phi) = {\case{\lambda}/{4!}} \phi^{4} ,
\end{equation}
one can derive the Feynman rules in configuration space :
\begin{enumerate}
	\item $G(\eta,\eta')$ for each boson line,
	\item $-\lambda /4!$ for each vertex,
	\item $ \ids$ for each internal vertex,
	\item standard symmetry factors,
\end{enumerate}
from which amplitudes may be calculated.

The determinant factor in Eq.(\ref{Sfgen}) deserves attention.  In flat space,
this
factor is usually absorbed into the normalization, but in this case it depends
on the radius of the four-sphere, $a$, and therefore the curvature of the
spacetime.  This factor contains the free field contribution of the scalar
field
to the trace anomaly.  Analogous factors appear for fields of different spin.

The generating functional for fermions may be defined using anticommuting
source terms\cite{ds1}
\begin{equation}
	Z_{0}[H,\bar{H}] = \int {\cal D} \psi \; {\cal D} \bar{\psi}
		\exp \left\{\ids \left[ -\bar{\psi} \gamma^{a}
                \left( D_{a} - \xonyasq{n}{2} \eta_{a} \right) \psi
		+ \bar{H} \psi + \bar{\psi} H \right] \right \} ,
\end{equation}
with the propagator given by
\begin{equation}
	S(\eta,\eta') = \left. \frac{\delta^{2}Z_{0}[H,\bar{H}]}{\delta \bar{H}(\eta)
			. - \delta H(\eta')} \right|_{H=\bar{H}=0}
			= \left[  \gamma \cdot \left( D - \xonyasq{n}{2}\eta \right)
			\right]^{-1} (\eta,\eta') .
		\label{spprop}
\end{equation}

Care is required for spin $1$ fields due to the presence of gauge and
radial degrees of freedom\cite{ds2}.  Consider the functional
\begin{equation}
	Z_{0}[J_{a}] = \int {\cal D} A \; \exp \left\{ \ids \left[
			\xonyasq{1}{2} A^{a} \pop{}{ab} A^{b} + J_{a} A^{a} \right] \right\} ,
			\label{Vfgen}
\end{equation}
due to the transformations of Eq.(\ref{gauge}) and Eq.(\ref{radial}) this is
ill defined and
therefore requires a gauge fixing term.  In order to motivate the choice of
such a term, note that the identities Eq.(\ref{P1}), Eq.(\ref{P2}) and
Eq.(\ref{P3})
imply that $A^{a}$ may be written in the form
\begin{equation}
	A^{a} = \pop{ab}{} B_{b} + D^{a}\Lambda + \onasq \eta^{a} \theta .
\end{equation}
If a gauge fixing operator is chosen of the form
\begin{equation}
	F^{ab} = \pop{ab}{} - N \kdelta{ab}{} ,
\end{equation}
then the result will vanish when acting on the transverse part $A_{\perp}^{a}
\equiv \pop{ab}{} B_{b}$,
suggesting the gauge condition
\begin{equation}
	F_{ab} A^{b} = 0 . \label{gfix}
\end{equation}
Following the standard procedure we insert the operator
\begin{equation}
	1 = \Delta(A) \int {\cal D}(\Lambda,\theta) \; \exp \left\{ \ids
		\left[ - \xonyasq{1}{2} \onal A'_{a} F^{ab} A'_{b} \right] \right \} ,
		\label{jacfix}
\end{equation}
into Eq.(\ref{Vfgen}) and factorize out the group integration.  The invariance
of the
Jacobian factor $\Delta(A)$ allows one to evaluate the integral in
Eq.(\ref{jacfix}) for $A^{a}$
satisfying the condition Eq.(\ref{gfix}) with the result
\begin{equation}
	\Delta(A)^{-1} = \left[ \det \onasq \left(- \hlsq \right) \right]^{-1}
			\left[ \det {\case{1}/{a^{4}}}\left( - N + n - 4 \right) \right]^{-\half} .
\end{equation}
The first factor is the usual Faddeev-Popov determinant while the second
comes about from the need to remove the radial degree of freedom.  The final
version of the generating function is
\begin{eqnarray}
	Z_{0}[J^{a}] & = & \left[ \det \onasq \left(-\hlsq \right) \right]
			\left[ \det {\case{1}/{a^{4}}}\left( - N + n - 4 \right)
			\right]^{\half} \nonumber \\
		& & \int {\cal D} A \; \exp \left\{ \ids
			\left[ \xonyasq{1}{2} A^{a} \left( \pop{}{ab} - \onal F_{ab} \right)
			A^{b} + J^{a} A_{a} \right] \right\} ,
\end{eqnarray}
with the progagator given by
\begin{equation}
	E^{ab}(\eta,\eta') = \left. \frac{\delta^{2} Z_{0}[J^{c}]}{\delta J_{a}(\eta)
				\delta J_{b}(\eta')} \right|_{J_{a}=0}
				= \left[ -\onasq \left( P - \onal F \right)
					\right]^{-1\, ab}(\eta,\eta').
\end{equation}
Putting this together,  one can write down the rules for spinor
electrodynamics
\begin{enumerate}
	\item $S_{\alpha \beta}(\eta,\eta')$ for each fermion line directed from
$\eta'$ to $\eta$,
	\item $E_{ab}(\eta,\eta')$ for each photon line connecting $\eta'$ to $\eta$,
	\item $ie \, \Q{\;\; b}{a}{} (\gamma_{b})_{\alpha \beta}$
			for each vertex with $\alpha$ outgoing and $\beta$ incoming,
	\item $\ids$ for each internal vertex,
	\item $(-1)$ for each fermion loop.
\end{enumerate}

\section{One-Particle Irreducible Amplitudes}

It is convienient in this approach to work with the $1$-particle
irreducible functions  which are generated by the effective action.  To do
this one must perform a Legendre transformation on the generating
function.  Considering the case of electrodynamics,  the effective action is
defined in Euclidean space by
\begin{equation}
	-\Gamma[A^{a},\bar{\psi},\psi] = G[J^{a},\bar{H},H]
		- \ids [J \cdot A + \bar{H} \phi + \bar{\phi} H],
\end{equation}
where
\begin{equation}
	Z[J^{a},\bar{H},H] = e^{G[J^{a},\bar{H},H]} ,
\end{equation}
\begin{equation}
	A_{a}(\eta) = \frac{\delta G}{\delta J^{a}(\eta)} , \; \;
	\bar{\psi}(\eta) = \frac{\delta G}{-H(\eta)} , \; \;
	\psi(\eta) = \frac{\delta G}{\delta \bar{H}(\eta)} ,
\end{equation}
and
\begin{equation}
        J_{a}(\eta) = \frac{\delta \Gamma}{\delta A^{a}(\eta)} , \; \;
        \bar{H}(\eta) = \frac{\delta \Gamma}{-\psi(\eta)} , \; \;
        H(\eta) = \frac{\delta \Gamma}{\delta \bar{\psi}(\eta)} .
\end{equation}
Since the generating function is invariant under a shift in the integration
variables one can use standard\cite{itzub} functional techniques to obtain the
equations of motion and the Ward Identities for the Greens functions of the
theory.  These may be expressed in terms of the irreducible amplitudes
through the Legendre transform.  For the inverse photon propagator one
obtains
\begin{equation}
	\Gamma_{ab}(\eta,\eta') = - \onasq \left( \pop{}{ab} - \onal F_{ab} \right)
				\delta(\eta - \eta') - \Pi_{ab}(\eta,\eta') ,
		\label{ph2}
\end{equation}
where $\Pi_{ab}(\eta,\eta')$ is the sum of all two point photon irreducible
diagrams with
external propagators  replaced by delta functions.  Invariance under gauge
and radial transformations leads to a set of Ward Identities
\begin{equation}
	\eta_{a} \left[ \frac{\delta \Gamma}{\delta A^{a}(\eta)} - \onasq \onal
F^{ab}A_{b} \right] = 0
		\label{Wrad}
\end{equation}
and
\begin{equation}
        D_{a} \left[ \frac{\delta \Gamma}{\delta A^{a}(\eta)} - \onasq \onal
F^{ab}A_{b} \right]
		= ie \left[ \frac{\delta \Gamma}{-\psi(\eta)} \psi - \bar{\psi}
		\frac{\delta \Gamma}{\delta \bar{\psi}(\eta)} \right],
		\label{Wgauge}
\end{equation}
Differentiating once with respect to $A_{b}(\eta')$ and setting sources to zero
gives
\begin{equation}
	\eta_{a} \Pi^{ab}(\eta,\eta') = D_{a} \Pi^{ab}(\eta,\eta') = 0 .
	\label{PiWardIds}
\end{equation}
These identities imply that $\Pi_{ab}(\eta,\eta')$ is of the form
\begin{equation}
	\Pi_{ab}(\eta,\eta') = \pop{}{ab} \Pi(\eta,\eta') ,
\end{equation}
which is the spherical analogue of the transversity condition in flat space.
Of course, these relations must be checked explicitly to ensure that no
anomalous quantum effects violate the identities and destroy the
renormalizability of the theory.

The fermion equation of motion similarly gives in terms of the inverse
propagator
\begin{equation}
	\Gamma_{\alpha \beta}(\eta,\eta') = \left[ \gamma \cdot \left(D -
\xonyasq{n}{2}\eta \right)
			\right] _{\alpha \beta} \delta(\eta - \eta')
			- \Sigma_{\alpha \beta}(\eta,\eta'),
		\label{sp2}
\end{equation}
where $\Sigma_{\alpha \beta}(\eta,\eta')$ is the sum of all irreducible, two
point fermion diagrams.

The irreducible vertex function is simply
\begin{equation}
	\Gamma^{a}_{\alpha \beta}(\eta_{1},\eta_{2},\eta)
		= - G^{a}_{\alpha \beta}(\eta_{1},\eta_{2},\eta)_{trunc.},
		\label{vt3}
\end{equation}
the negative of the truncated three point function.  The Ward Identities
Eq.(\ref{Wrad}) and Eq.(\ref{Wgauge}) then imply
\begin{equation}
	\eta_{a} \Gamma^{a}_{\alpha \beta}(\eta_{1},\eta_{2},\eta) = 0 ,
	\label{VWIone}
\end{equation}
and
\begin{equation}
	D_{a} \Gamma^{a}_{\alpha \beta}(\eta_{1},\eta_{2},\eta) = ie
		\left[ \Gamma_{\alpha \beta}(\eta_{1},\eta) \delta(\eta - \eta_{2}) -
			\Gamma_{\alpha \beta}(\eta,\eta_{2}) \delta(\eta - \eta_{1}) \right].
	\label{VWItwo}
\end{equation}
For the theory to be renormalizable, divergent parts of the amplitudes
Eq.(\ref{ph2}), Eq.(\ref{sp2}) and Eq.(\ref{vt3}) must be of the same form as
the tree level functions.
An amplitude must therefore be treated as a distribution and expanded in a
series of operators acting on delta functions\cite{drum2}.  First consider the
vacuum
polarization Eq.(\ref{ph2}), write
\begin{equation}
	\Pi^{ab}(\eta,\eta') = \left[ A_{0}\Q{ab}{}{} + A_{1}\pop{ab}{}(\eta)
			+ A_{2}F^{ab} + \ldots \right ] \delta(\eta - \eta'),
	\label{Pidist1}
\end{equation}
where we have allowed for the possibility of a mass term $A_{0}$ and a gauge
fixing part $A_{2}$, which should be zero by the Ward Identities.  One
can, by various means, extract the values of the expansion coefficients.
Contracting the indices gives
\begin{equation}
	\Pi^{a}\, _{a}(\eta,\eta') = \left[ A_{0}n + A_{1}(n-1)(\hlsq)
		+ A_{2} \left[ (n+1)(n-2) - 2(\hlsq)\right] + \ldots \right] \delta(\eta -
\eta'),
	\label{Pidist2}
\end{equation}
and so
\begin{equation}
	\idssp  \Pi^{a}\, _{a}(\eta,\eta') =
		\vol  \left[  A_{0}n + A_{2}(n+1)(n-2) \right] ,
	\label{Pidist3}
\end{equation}
with the derivative terms vanishing due to integration by parts.  Similarly
one can show that
\begin{equation}
	\idssp \xonyasq{1}{2} (\eta - \eta')^{2} \Pi^{a}\, _{a}(\eta,\eta') =
		 \vol  \left[ A_{1}n(n-1) - 2A_{2}n \right]
	\label{Pidist4}
\end{equation}
and
\begin{equation}
	\idssp \onasq \eta_{a} \eta_{b} \Pi^{ab}(\eta, \eta') = \vol  A_{2}(n-2).
	\label{Pidist5}
\end{equation}

For the fermion self energy Eq.(\ref{sp2}), $\Sigma_{\alpha \beta}(\eta,\eta')$
may be expanded in the
form
\begin{eqnarray}
	\Sigma_{\alpha \beta}(\eta,\eta') & = & \left[ B_{0}(I) + \onasq
B_{1}(\gamdeta{})
		+ B_{2} \gamma \cdot (D - \xonyasq{n}{2}\eta) \right. \nonumber \\
		& & \indb \left.
		\mbox{} + B_{3} (\gamdeta{})(\gamdD{}) \ldots \right]_{\alpha \beta}
		\delta(\eta - \eta')  ,
	\label{Sigd1}
\end{eqnarray}
where the possiblity of mass  and chiral terms has been considered.  However,
the only  coefficient allowed to be divergent is $B_{2}$.  To extract these
constants
one requires a set of gamma matrix identities and traces, these are given in
Appendix(\ref{gammaid}).  For reasons of notational clarity, traces over spinor
indices will
be denoted by $TR\{ \ldots \}$ .  Using the results of Appendix(\ref{gammaid})
and integrating by
parts when necessary one obtains
\begin{equation}
	\idssp TR\left\{ \Sigma(\eta,\eta') \right\} = \vol 2^{n/2} B_{0} ,
	\label{Sigd2}
\end{equation}
\begin{equation}
        \idssp TR\left\{ (\gamdeta{})\Sigma(\eta,\eta') \right\} = \vol 2^{n/2}
		(B_{1} - \nontwo B_{2}),
	\label{Sigd3}
\end{equation}
\begin{equation}
        \idssp TR\left\{ (\gamdeta{'})\Sigma(\eta,\eta') \right\} = \vol
2^{n/2}
		(B_{1} + \nontwo B_{2}),
	\label{Sigd4}
\end{equation}
and
\begin{equation}
        \idssp TR\left\{ (\gamdeta{})(\gamdeta{'})\Sigma(\eta,\eta') \right\} =
		\vol 2^{n/2} (B_{0} + nB_{3}).
	\label{Sigd5}
\end{equation}

The vertex function may be expanded in the same way and is left as an
exercise.

\section{Transform Space Representaton}

We have arrived at the central problem, the need for a consistent method
of evaluating integrals of the form
\begin{equation}
	X = \int d\sigma_{1} \: d\sigma_{2} \ldots d\sigma_{N}
		f( \eta_{1}, \eta_{2} , \ldots , \eta_{N} ) ,
\end{equation}
where $f( \eta_{1}, \eta_{2} , \ldots , \eta_{N} )$
is a scalar function which may contain factors of $\eta_{i}$,
propagators, vertex functions and derivatives.  Unfortunately the
configuration space methods of Drummond\cite{drum2} fall down here because the
photon propagator cannot be written in the form $\left| \eta - \eta'
\right|^{m}$ in a general
gauge.  Also there is great difficulty when indices contract between integrals
over different
coordinates particularly when these involve derivative couplings.
The solution is to go to angular momentum space and expand
propagators in terms of spherical harmonics.  The method here is applicable
provided there
are no more than two harmonics at the same point.  However, this problem does
not arise for the
one-loop diagrams we consider here.

Consider the case of the scalar field.  The appropriate basis functions
are the scalar harmonics satisfying
\begin{equation}
	(\hlsq) \schm{}{}{} = -l(l+n-1) \schm{}{}{},
	\label{sceigen}
\end{equation}
where $m = 1 \ldots dim(l,0)$, the dimension of the representation, given by
\begin{equation}
	dim(l,0) = \dlz .
\end{equation}
Here we use $(\mu_{1},\mu_{2})$ to label the represenation of $S0(5)$ in common
with Refs.\cite{ds2,ds3}.
The harmonics satisfy the orthonormaliy and completeness conditions
\begin{equation}
	\ids \cschm{'}{'}{} \schm{}{}{} = \delta_{l,l'} \delta_{m,m'}
\end{equation}
and
\begin{equation}
	\sum_{lm} \schm{}{}{} \cschm{}{}{'} = \delta(\eta - \eta').
	\label{sccomp}
\end{equation}
Refering to Eq.(\ref{scprop}), the operator $M$ is diagonal
in $l$-space and so using Eq.(\ref{sceigen}) and Eq.(\ref{sccomp}) one obtains
\begin{equation}
	G(\eta,\eta') = \sum_{lm} \frac{a^{2}}{(l+\nontwo)(l+\nontwo-1)}
	\schm{}{}{} \cschm{}{}{'} ,
\end{equation}
or
\begin{equation}
	G(\eta,\eta') \equiv a^{2} \sum_{lm} G(l) \schm{}{}{} \cschm{}{}{'} .
\end{equation}

For the spinor field one requires a corresponding set of spinor
harmonics.  The tensor product of the spin $\half$
representation and scalar harmonics, $(\half,\half) \otimes (l,0) =
(l+\half,\half) \oplus
(l-\half,\half)$ gives two inequivalent representations each with same
$(\hlsq)$
eigenvalue classified by their eigenvalue with respect to $(\sigdl)$, where
\begin{equation}
	\Sigma_{ab} = {\case{1}/{4}} \left[ \gamma_{a}, \gamma_{b} \right] ,
\end{equation}
are the spin $\half$ rotation generators.  We have
\begin{equation}
	(\hlsq) \sphm{}{}{}{\mu}{} = -l(l+n-1) \sphm{}{}{}{\mu}{} ,
	\label{speigen1}
\end{equation}
where the label $s$ spans the representation and the label $(\mu)$, defined to
be
$+\half$ or $-\half$, refers to the $(l+\half,\half)$ or the $(l-\half,\half)$
representations
respectively.  The $(\sigdl)$ eigenvalues are
\begin{equation}
	(\sigdl) \sphm{}{}{}{+\half}{} = -l \sphm{}{}{}{+\half}{} ,
	\label{speigen2}
\end{equation}
and
\begin{equation}
	 (\sigdl) \sphm{}{}{}{-\half}{} = (l+n-1) \sphm{}{}{}{-\half}{} .
	\label{speigen3}
\end{equation}
The orthonormality and completeness conditions are
\begin{equation}
	\ids \csphm{'}{'}{}{\mu'}{\alpha} \sphm{}{}{}{\mu}{\alpha} = \delta_{l',l}
\delta_{s,s'}
		\delta_{\mu,\mu'} ,
	\label{sporth}
\end{equation}
and
\begin{equation}
	\sum_{l \mu s} \sphm{}{}{}{\mu}{\alpha} \csphm{}{}{'}{\mu}{\beta} =
		\delta_{\alpha \beta} \delta(\eta - \eta') .
	\label{spcomp}
\end{equation}
Keeping in mind that $(\gamdeta{})(\sigdl) = a^{2}(\gamdD{})$,
Eq.(\ref{spprop}),
Eq.(\ref{speigen1})\ldots (\ref{spcomp}) and a little algebra give
\begin{eqnarray}
	S_{\alpha \beta}(\eta,\eta') & = (\gamdeta{})_{\alpha \delta} {\displaystyle
\sum_{ls}} & \left[
		\frac{1}{(l+\case{n}/{2})} \sphm{}{}{}{+\half}{\delta}
		\csphm{}{}{'}{+\half}{\beta} \right. \nonumber \\
		& & \left. \mbox{} \; \;  - \frac{1}{(l+\case{n}/{2}-1)}
		\sphm{}{}{}{-\half}{\delta} \csphm{}{}{'}{-\half}{\beta} \right]
	\label{splprop1}
\end{eqnarray}
which can be more compactly written as
\begin{eqnarray}
	 S_{\alpha \beta}(\eta,\eta')
		& \equiv & (\gamdeta{})_{\alpha \delta} {\displaystyle \sum_{l\mu s}}
		S(l,\mu) \sphm{}{}{}{\mu}{\delta} \csphm{}{}{'}{\mu}{\beta} \\
		& \equiv  & - {\displaystyle \sum_{l\mu s}}
		S(l,\mu) \sphm{}{}{}{\mu}{\alpha} \csphm{}{}{'}{\mu}{\delta} (\gamdeta{'})
		_{\delta \beta} .
	\label{splprop2}
\end{eqnarray}

For the vector case we get for the tensor product $(l,0) \otimes (1,0) =
(l-1,0) \oplus
(l,1) \oplus (l+1,0)$, three inequivalent representations which we will denote
by $\lambda =
-1,0,+1$ respectively.  The dimension of the representation $(l,1)$ is
\begin{equation}
	dim(l,1) = \dlone .
\end{equation}
Then
\begin{equation}
	(\hlsq) \vhm{}{}{}{\lambda}{a}{} = -l(l+n-1) \vhm{}{}{}{\lambda}{a}{},
	\label{veigen1}
\end{equation}
and each is and eigenfunction of $(\half  L \cdot S)_{ab} = \lop{}{ab}$ , where
\begin{equation}
	(S_{ab})_{cd} = \kdelta{}{ac} \kdelta{}{bd} - \kdelta{}{ad} \kdelta{}{bc} ,
\end{equation}
are the spin $1$ rotation generators.  The eigenvalues are
\begin{eqnarray}
	\lop{}{ab} \vhm{}{}{}{+1}{b}{} & = &  -l \vhm{}{}{}{+1}{}{a} , \label{veigen2}
\\
	\lop{}{ab} \vhm{}{}{}{0}{b}{} & = & 1 \vhm{}{}{}{0}{}{a} , \label{veigen3} \\
	\lop{}{ab} \vhm{}{}{}{-1}{b}{} & = & (l+n-1) \vhm{}{}{}{-1}{}{a}
\label{veigen4} .
\end{eqnarray}
The orthonormality and completeness conditions are
\begin{equation}
	\ids \cvhm{}{}{}{\lambda}{}{a} \vhm{'}{'}{}{\lambda'}{a}{} =
		\delta_{l,l'} \delta_{m,m'} \delta_{\lambda,\lambda'} ,
	\label{vorth}
\end{equation}
and
\begin{equation}
	\sum_{l\lambda m} \vhm{}{}{}{\lambda}{}{a} \cvhm{}{}{'}{\lambda}{}{b}
		= \kdelta{}{ab} \delta(\eta - \eta') .
	\label{vcomp}
\end{equation}
Clearly this procedure may be carried out for higher spin representations
such as spin $\case{3}/{2}$ and spin $2$.

The kinetic operators $\pop{}{ab}$ and $F_{ab}$ for the vector field are
diagonalized
in this representation.  $\pop{}{ab}$ has zero eigenvalue for $\lambda = -1$ or
$+1$ and
$-(l+1)(l+n-2)$ for $\lambda = 0$.  $F_{ab}$ has zero eigenvalue for $\lambda =
0$
and $(l+1)(l+n-2)$ for $\lambda = -1$ or $+1$ .
This illustrates the utility of the choice of gauge, Eq.(\ref{gfix}), and leads
to
the convenient form of the propagator
\begin{eqnarray}
	E_{ab}(\eta,\eta') & = & \alpha {\displaystyle \sum_{l\lambda m}}
		\frac{a^{2}}{(l+1)(l+n-2)}
		\vhm{}{}{}{\lambda}{}{a} \cvhm{}{}{'}{\lambda}{}{b} \nonumber \\
		& & \mbox{} + (1-\alpha) {\displaystyle \sum_{lm}}
		\frac{a^{2}}{(l+1)(l+n-2)}
		\vhm{}{}{}{0}{}{a} \cvhm{}{}{'}{0}{}{b} ,
	\label{Vlprop1}
\end{eqnarray}
which again may be compactly written
\begin{eqnarray}
	 E_{ab}(\eta,\eta') & = & \alpha\, a^{2}\, {\displaystyle \sum_{l\lambda m}}
		E(l) \vhm{}{}{}{\lambda}{}{a} \cvhm{}{}{'}{\lambda}{}{b} \nonumber \\
		& & \mbox{} + (1-\alpha)\, a^{2}\, {\displaystyle \sum_{lm}}
		E(l) \vhm{}{}{}{0}{}{a} \cvhm{}{}{'}{0}{}{b} .
	\label{Vlprop2}
\end{eqnarray}

\section{Operator Matrix Elements}

We define the matrix element of an operator between two scalar
harmonics as
\begin{equation}
	\left( O_{l',l}^{(\cdots)} \right)_{m',m} \equiv
		\ids \cschm{'}{'}{} O^{(\cdots)} (\eta) \schm{}{}{} ,
\end{equation}
where $(\cdots)$ refers to any indices carried by the operator.  The simplest
example is that of $\lop{}{ab}$ , which commutes with $(\hlsq)$, we define
\begin{equation}
	\tlm{}{ab}{}{'}{} \delta_{l,l'} =
		\ids \cschm{'}{'}{} \lop{ab}{} \schm{}{}{} .
\end{equation}
The matrix $\tlm{}{ab}{}{'}{} \delta_{l,l'}$
is a square $dim(l,0) \otimes dim(l,0)$ matrix which satisfies the
same commutation relations as $\lop{ab}{}$ .  A list of useful commutation and
other
relations for both operators and matrix elements is given in
Appendix(\ref{OpId}).   The
next two operators which we must deal with are $D_{a}$ and $\eta_{a}$ .  Using
the
commutation relations
\begin{equation}
	\left[ \hlsq, \onasq \eta_{a} \right] = - \xonyasq{n}{} \eta_{a} + 2D_{a}
\end{equation}
and
\begin{equation}
	\left[ \hlsq , D_{a} \right] = - \xonyasq{2}{} \eta_{a}(\hlsq) + (n-2)D_{a} ,
\end{equation}
one is able to show that
\begin{equation}
	\hlsq \left( D_{a} + \onasq l\, \eta_{a} \right) \schm{}{}{} =
		-(l-1)(l+n-2)\left( D_{a} + \onasq l\, \eta_{a} \right) \schm{}{}{}
\end{equation}
and
\begin{equation}
	\hlsq \left( D_{a} - \onasq(l+n-1) \eta_{a} \right) \schm{}{}{} =
		-(l+1)(l+n)  \left( D_{a} - \onasq(l+n-1) \eta_{a} \right) \schm{}{}{} .
\end{equation}
The completeness relation Eq.(\ref{sccomp}) implies that
\begin{equation}
	\left( D_{a} + \onasq l \, \eta_{a} \right) \schm{}{}{} \equiv
		\sum_{m'=1}^{dim(l-1,0)} \vlm{-1}{}{a}{'}{} \schm{-1,}{'}{}
	\label{Vlmdef}
\end{equation}
and
\begin{equation}
	\left( D_{a} - \onasq(l+n-1) \eta_{a} \right) \schm{}{}{} \equiv
		\sum_{m'=1}^{dim(l+1,0)} \ulm{+1}{}{a}{'}{} \schm{+1,}{'}{} .
	\label{Ulmdef}
\end{equation}
The matrix $\vlm{}{}{a}{'}{}$ is $dim(l,0) \otimes dim(l+1,0)$ and
$\ulm{}{}{a}{'}{}$ is $dim(l,0) \otimes dim(l-1,0)$.  By inverting the above
one obtains
\begin{eqnarray}
	\ids \cschm{'}{'}{} \onasq \eta_{a} \schm{}{}{} & =
		{\displaystyle \frac{1}{(2l+n-1)}}
		& \left[ \vlm{-1}{}{a}{'}{} \delta_{l',l-1} \right. \nonumber \\
		& & \left. \mbox{} \; \; - \ulm{+1}{}{a}{'}{} \delta_{l',l+1} \right]
	\label{Etadef}
\end{eqnarray}
and
\begin{eqnarray}
	\ids \cschm{'}{'}{} D_{a} \schm{}{}{} & =
		{\displaystyle \frac{1}{(2l+n-1)}}
		& \left[ (l+n-1)\vlm{-1}{}{a}{'}{} \delta_{l',l-1} \right. \nonumber \\
		& & \left. \mbox{} \; \; + l\,  \ulm{+1}{}{a}{'}{} \delta_{l',l+1} \right] .
	\label{Ddef}
\end{eqnarray}
By further application of the operators in Eq.(\ref{Vlmdef}) and
Eq.(\ref{Ulmdef}) it can be shown that
\begin{equation}
	\sum_{m''=0}^{dim(l+1,0)} \vlm{}{}{a}{}{''} \vlm{+1}{a}{}{''}{'} =
	\sum_{m''=0}^{dim(l-1,0)} \ulm{}{}{a}{}{''} \ulm{-1}{a}{}{''}{'} = 0 ,
\end{equation}
\begin{equation}
	\sum_{m''=0}^{dim(l+1,0)} \vlm{}{}{a}{}{''} \ulm{+1}{a}{}{''}{'} =
		- \onasq (l+n-1)(2l+n+1) \delta_{m,m'}
\end{equation}
and
\begin{equation}
	\sum_{m''=0}^{dim(l-1,0)} \ulm{}{}{a}{}{''} \vlm{-1}{a}{}{''}{'} =
		- \onasq l\, (2l+n-3) \delta_{m.m'} .
\end{equation}
Integrating by parts leads to the conjugate relations
\begin{eqnarray}
	\ulm{}{}{a}{'}{}^{\ast} & = & \ids \left( D_{a} - \onasq(l+n-2) \eta_{a}
\right)
		\cschm{-1,}{}{} \schm{}{'}{} \nonumber \\
	& = & {\displaystyle \frac{(2l+n-3)}{(2l+n-1)}} \vlm{-1}{a}{}{}{'}
\end{eqnarray}
and similarly
\begin{equation}
	\vlm{}{}{a}{'}{}^{\ast} = - \frac{(2l+n+1)}{(2l+n-1)} \ulm{+1}{}{a}{}{'} .
\end{equation}
Since these matrices may be non-square one
must be careful to ensure that the multiplication is valid.  Note that with
these conventions, the $l$ label implies the dimensionality of the left index.

In order to easily write down the matrix elements of the the commonly used
operators
$D_{a}$, $\eta_{a}$ and $\Qop{}{ab}$ which do not commute with $\hlsq$ some
more notation
is useful. Define a new index $(\sigma)$ which indicates the shift in $l$
eigenvalue
between the left and right $m$ indices.  Therefore for some operator
$O^{(\cdots)}(\eta)$ we have
\begin{equation}
	\left( O_{l}^{(\sigma)(\cdots)} \right)_{mm'} \equiv
		\ids \cschm{}{}{} O^{(\cdots)}(\eta) \schm{+\sigma \,}{'}{} .
\end{equation}
In particular, Eq.(\ref{Etadef}) and Eq.(\ref{Ddef}) imply that
\begin{equation}
	\etal{}{+1}{}{a}_{mm'} = + a^{2} \frac{1}{(2l+n+1)} \vlm{}{}{a}{}{'} ,
\end{equation}
\begin{equation}
	\etal{}{-1}{}{a}_{mm'} = - a^{2} \frac{1}{(2l+n-3)} \ulm{}{}{a}{}{'} ,
\end{equation}
\begin{equation}
	\Dl{}{+1}{}{a}_{mm'} = \frac{(l+n)}{(2l+n+1)} \vlm{}{}{a}{}{'}
\end{equation}
and
\begin{equation}
	\Dl{}{-1}{}{a}_{mm'} = \frac{(l-1)}{(2l+n-3)} \ulm{}{}{a}{}{'} .
\end{equation}
Henceforth, the $m$ indices will be suppressed with the normal rules of
matrix multiplication applying.
Refering to Eq.(\ref{Qdef}) the above allows one to write down the matrix
elements of $\Q{}{ab}{}$
\begin{eqnarray}
	\Ql{}{+2}{}{ab}	& = & - \onasq \etal{}{+1}{}{a} \etal{+1}{+1}{}{b} \nonumber
\\
		& = & - {\displaystyle \frac{a^{2}}{(2l+n+1)(2l+n+3)}}
			\vl{}{}{a} \vl{+1}{}{b} ,
	\label{Qlplus}
\end{eqnarray}
\begin{eqnarray}
	\Ql{}{0}{}{ab} & = & \kdelta{}{ab} - \onasq \etal{}{+1}{}{a}
\etal{+1}{-1}{}{b}
				- \onasq \etal{}{-1}{}{a} \etal{-1}{+1}{}{b} \nonumber \\
		& = & \kdelta{}{ab} +
		{\displaystyle \frac{a^{2}}{(2l+n+1)(2l+n-1)}} \vl{}{}{a} \ul{+1}{}{b}
\nonumber \\
		& & \mbox{} + {\displaystyle \frac{a^{2}}{(2l+n-3)(2l+n-1)}} \ul{}{}{a}
\vl{-1}{}{b}
	\label{Qlzero}
\end{eqnarray}
and
\begin{eqnarray}
	\Ql{}{-2}{}{ab} & = & - \onasq \etal{}{-1}{}{a} \etal{-1}{-1}{}{b} \nonumber
\\
                & = & - {\displaystyle \frac{a^{2}}{(2l+n-3)(2l+n-5)}}
                        \ul{}{}{a} \ul{-1}{}{b} .
	\label{Qlminus}
\end{eqnarray}

\section{Matrix Elements of Harmonics and The Projection Operator Theorem}

Acting on the left of Eq.(\ref{Vlmdef}) and Eq.(\ref{Ulmdef}) with $\lop{}{ab}$
leads to the
eigenvalue relations
\begin{equation}
	\tl{}{}{ab} \vl{}{b}{} = -l \vl{}{}{a} ,
	\label{Veigen}
\end{equation}
and
\begin{equation}
	\tl{}{}{ab} \ul{}{b}{} = (l+n-1) \ul{}{}{a} .
	\label{Ueigen}
\end{equation}
The eigenvalue relations Eq.(\ref{Veigen}) and Eq.(\ref{Ueigen}) imply a
connection with the vector harmonics and it can be shown that
\begin{equation}
	\vhm{}{}{}{+1}{}{a} = N^{+}(l) \sum^{dim(l,0)}_{m'} \vlm{}{}{a}{'}{}
\schm{}{'}{} ,
	\label{Vplus}
\end{equation}
\begin{equation}
	\vhm{}{}{}{-1}{}{a} = N^{-}(l) \sum^{dim(l,0)}_{m'} \ulm{}{}{a}{'}{}
\schm{}{'}{} ,
	\label{Vminus}
\end{equation}
where the normalization factors are
\begin{equation}
	N^{+}(l) = \frac{a}{\sqrt{(l+1)(2l+n+1)}} ,
\end{equation}
\begin{equation}
	N^{-}(l) = \frac{a}{\sqrt{(l+n-2)(2l+n-3)}} ,
\end{equation}
give a representation of the longitudinal vector harmonics.
We define the matrix element of the transverse harmonic
\begin{equation}
	\wlm{}{}{a}{}{} = \ids \cschm{}{}{} \tvhm{}{}{}{}{a} ,
	\label{Vzero}
\end{equation}
where the upper case $M$ is used to indicate that the dimension is $dim(l,1)$
instead of $dim(l,0)$.  The eigenvalue condition Eq.(\ref{veigen3}) gives
\begin{equation}
	\tl{}{}{ab} \wl{}{b}{} = 1 \wl{}{}{a} ,
\end{equation}
while the orthonormality condition Eq.(\ref{vorth}) implies
\begin{eqnarray}
	\vl{-1}{}{a} \wl{}{a}{} & = & 0 , \\
	\ul{+1}{}{a} \wl{}{a}{} & = & 0 , \\
	\wl{}{a}{}^{\ast} \wl{}{a}{} & = & 1 .
\end{eqnarray}
Note that since $\wlm{}{}{a}{}{}$ cannot be written as the matrix element of a
local
operator we have no way to relate it to its conjugate $\cwlm{}{}{a}{}{}$ .

For the spinor harmonics we define the elements
\begin{equation}
	\chilms{}{\mu}{\alpha}{}{} = \ids \cschm{}{}{} \sphm{}{}{}{\mu}{\alpha} ,
\end{equation}
and the eigenvalue conditions Eq.(\ref{speigen2}) and Eq.(\ref{speigen3}) imply
\begin{eqnarray}
	\sigdtl{} \chil{}{+\half} & = &  -l \chil{}{+\half} , \\
	\sigdtl{} \chil{}{-\half} & = & (l+n-1) \chil{}{-\half} .
\end{eqnarray}
Again we have the case where $\chilms{}{\mu}{}{}{}$ may not be written as the
element of
a local operator.

In order for these definitions to be useful we require something more.
The key is that propagators contain harmonics in conjugate pairs allowing
us to appeal to the completeness conditions.  Consider a complete set of
harmonics in some spin $S$ representation of $SO(n+1)$.  The completeness
relations for these and the scalar harmonics imply that
\begin{equation}
	\sum_{\lambda m'} \vhm{}{'}{}{\lambda}{}{[X]} \cvhm{}{'}{'}{\lambda}{}{[Y]}
		= \kdelta{}{[X],[Y]} \sum_{m} \schm{}{}{} \cschm{}{}{'} ,
	\label{Scomp}
\end{equation}
for each value of $l$, where $[X]$ refers to whatever indices are carried by
the
representation.  Now the spin $S$ representation is split into irreducible
parts
denoted by the label $\lambda$ each with a different eigenvalue of $(L \cdot
S)$ where the
$S_{[X][Y]}$  operators are the spin $S$ generators.  By virtue of the
eigenvalue
conditions, we can form projection operators which project onto the
subspace of a particular irreducible representation, we call this operator
$\vp{}{\lambda}{}{[X],[Y]}{}$.  Acting on the left of Eq.(\ref{Scomp}) produces
\begin{equation}
	\sum_{m'} \vhm{}{'}{}{\lambda}{}{[X]} \cvhm{}{'}{'}{\lambda}{}{[Y]}
                = \sum_{m} \vp{}{\lambda}{}{[X],[Y]}{} \schm{}{}{}
\cschm{}{}{'} .
\end{equation}
By multiplying by $\schm{}{'}{'}$, $\cschm{}{}{}$ and integrating over $\eta$
and $\eta'$ we obtain
the final form of the Projection Operator Theorem
\begin{equation}
	\vl{}{(\lambda)}{[X]} \vl{}{(\lambda)}{[Y]}^{\ast} =
\vpl{}{\lambda}{}{[X],[Y]} ,
\end{equation}
where the matrices $\vpl{}{\lambda}{}{[X],[Y]}$
 are known combinations of $\tl{}{}{ab}$ and $S_{[X][Y]}$ .

For the case of spin $\half$, using Eq.(\ref{speigen2}) and
Eq.(\ref{speigen3}),
one obtains
\begin{eqnarray}
	\p{}{+\half}{} & = & \frac{1}{(2l+n-1)} \left[ (l+n-1) - (\sigdl) \right] ,
\nonumber \\
	\p{}{-\half}{} & = & \frac{1}{(2l+n-1)} \left[ l + (\sigdl) \right],
	\label{spinpl}
\end{eqnarray}
and so
\begin{eqnarray}
	\chil{}{+\half} \chil{}{+\half}^{\ast} & = &  \frac{1}{(2l+n-1)} \left[
(l+n-1) - \sigdtl{} \right] ,
	\nonumber \\
	\chil{}{-\half} \chil{}{-\half}^{\ast} & = & \frac{1}{(2l+n-1)} \left[ l +
\sigdtl{} \right] .
	\label{spinplm}
\end{eqnarray}
Refering to Eq.(\ref{splprop1}), Eq.(\ref{splprop2}) and Eq.(\ref{spinpl}) one
gets
\begin{equation}
	S(\eta,\eta') = (\gamdeta{}) \sum_{l\mu} S(l,\mu) \p{}{\mu}{}  \sum_{m}
\schm{}{}{} \cschm{}{}{'}
\end{equation}
but it is easy to show that
\begin{equation}
	\sum_{\mu} S(l,\mu) \p{}{\mu}{} = \spop{} G(l)
\end{equation}
where $G(l)$ is the $l$-space representation of the scalar propagator.  This
leads to the convienient
form of the spinor propagator
\begin{eqnarray}
	S(\eta,\eta') & = & \gamma \cdot \left[ \onasq (\Aterm) \eta - D \right]
G(\eta,\eta')
		\nonumber \\
		& = & (\gamdeta{}) \sum_{lm} G(l) \spop{} \schm{}{}{} \cschm{}{}{'}.
\end{eqnarray}

For the vector case, the eigenvalue relations Eq.(\ref{veigen2}) \ldots
Eq.(\ref{veigen4}) allow
one to write
\begin{eqnarray}
	\vp{}{+1}{}{ab}{} & = & \frac{1}{(l+1)(2l+n-1)} \left[ \lop{}{ac} -
\kdelta{}{ac} \right]
				\left[ \lop{c}{\;\; b} - (l+n-1) \kdelta{c}{\;\; b} \right] \nonumber \\
		& = & \frac{1}{(l+1)(2l+n-1)} \left[ \lop{}{ac}\lop{c}{\;\; b} -
(l+n)\lop{}{ab}
				+(l+n-1)\kdelta{}{ab} \right] \\
	\vp{}{0}{}{ab}{} & = & - \frac{1}{(l+1)(l+n-2)} \left[ \lop{}{ac} + l
\kdelta{}{ac} \right]
				\left[ \lop{c}{\;\; b} - (l+n-1) \kdelta{c}{\;\; b} \right] \nonumber \\
		& = & - \frac{1}{(l+1)(l+n-2)} \left[ \lop{}{ac}\lop{c}{\;\; b} -
(n-1)\lop{}{ab}
				-l(l+n-1)\kdelta{}{ab} \right] \\
	\vp{}{-1}{}{ab}{} & = & \frac{1}{(l+n-2)(2l+n-1)} \left[ \lop{}{ac} -
\kdelta{}{ac} \right]
				\left[ \lop{c}{\;\; b} + l \kdelta{c}{\;\; b} \right] \nonumber \\
		& = & \frac{1}{(l+n-2)(2l+n-1)} \left[ \lop{}{ac}\lop{c}{\;\; b} +
(l-1)\lop{}{ab}
				-l\kdelta{}{ab} \right] .
\end{eqnarray}
Note the relation between $\vp{}{0}{}{ab}{}$ and the operator $\pop{}{ab}$
introduced earlier.  Using
Eq.(\ref{Vplus}), Eq.(\ref{Vminus}) and Eq.(\ref{Vzero}) together with the
conjugate forms gives
\begin{equation}
	\vpl{}{+1}{}{ab} = \left[ N^{+}(l) \right]^{2} \vl{}{}{a} \vl{}{}{b}^{\ast}
				= - \frac{a^{2}}{(l+1)(2l+n-1)} \vl{}{}{a} \ul{+1}{}{b} ,
	\label{Vplplus}
\end{equation}
\begin{equation}
	\vpl{}{0}{}{ab} = \wl{}{}{a} \wl{}{}{b}^{\ast} ,
	\label{Wplzero}
\end{equation}
\begin{equation}
	\vpl{}{-1}{}{ab} = \left[ N^{-}(l) \right]^{2} \ul{}{}{a} \ul{}{}{b}^{\ast}
				= - \frac{a^{2}}{(l+n-2)(2l+n-1)} \ul{}{}{a} \vl{-1}{}{b} .
	\label{Uplminus}
\end{equation}
The vector propagator Eq.(\ref{Vlprop2}) may be rewritten using the
$\vp{}{\lambda}{}{ab}{}$ operators
\begin{eqnarray}
	E_{ab}(\eta,\eta') & = & \alpha\, a^{2}\, \sum_{l} E(l) \sum_{\lambda}
\vp{}{\lambda}{}{ab}{}
			\sum_{m} \schm{}{}{} \cschm{}{}{'} \nonumber \\
		& & \mbox{} + (1-\alpha)\, a^{2}\, \sum_{l} E(l) \vp{}{0}{}{ab}{}
				\schm{}{}{} \cschm{}{}{'} \\
		& = & \alpha\, a^{2} \: \kdelta{}{ab} \sum_{lm} E(l) \schm{}{}{}
\cschm{}{}{'} \nonumber \\
		& & \mbox{} + (1-\alpha)\, a^{2}\, \sum_{lm} E(l) \vp{}{0}{}{ab}{}
				\schm{}{}{} \cschm{}{}{'} .
\end{eqnarray}

Since the action of projection operators can considerably simplify calculations
it is useful to
rewrite Eq.(\ref{Qlzero}) using vector projection operators.  Using the fact
that
$\kdelta{}{ab} = \sum_{\lambda} \vpl{}{\lambda}{}{ab}$ and Eq.(\ref{Vplplus})
\ldots Eq.(\ref{Uplminus})
one obtains
\begin{equation}
	\Ql{}{0}{}{ab} = \vpl{}{0}{}{ab} + \frac{(l+n)}{(2l+n+1)} \vpl{}{+1}{}{ab}
				+ \frac{(l-1)}{(2l+n-3)} \vpl{}{-1}{}{ab} .
	\label{Qlzeropl}
\end{equation}

\section{One Loop Renormalisation of Spinor QED}

To see how all this works in practice we will compute the divergent parts of
the
vacuum polarization, spinor self-energy and the vertex function for spinor QED.

\subsection{Vacuum Polarization}

Using the rules in Section.(\ref{Quantization}) one can write down the
ampiltude to the
vacuum polarization (FIG.\ref{vacfig})
\begin{equation}
	\Pi_{ab}(\eta,\eta') = (-1) (ie)^{2} (\mu)^{4-n} TR \left\{
				\Q{}{ac}{} \gamma^{c} S(\eta,\eta') \gamma_{d} \Q{d}{\;\; b}{'}
				S(\eta',\eta) \right\} .
	\label{Pione}
\end{equation}
In order to expand $\Pi_{ab}(\eta,\eta')$ as a distribution we refer to
Eq.(\ref{Pidist1})
\ldots Eq.(\ref{Pidist5}) .  The contraction of Eq.(\ref{Pione}) with $\onasq
\eta^{a} \eta^{b}$
is zero due to the $Q$ operators, implying that $A_{2}$ vanishes identically.
Therefore, using Eq.(\ref{Pidist3}) one has
\begin{equation}
	A_{0} = \FactorOne \idssp \Q{a}{\;\; c}{} \Q{d}{\;\; a}{'}
		TR \left\{ \gamma^{c} S(\eta,\eta') \gamma_{d} S(\eta',\eta) \right\} .
\end{equation}
Putting in the $l$-space form of the propagators and writing $(\gamdeta{})
\equiv
\slsh{\eta}$, this becomes
\begin{eqnarray}
	A_{0} & = & \FactorOne \sum_{lm} \sum_{l'm'} G(l) G(l')
			\idssp \Q{a}{\;\; c}{} \Q{d}{\;\; a}{'} \nonumber \\
		& & .-TR\left\{ \gamma^{c} \spop{} \schm{}{}{} \cschm{}{}{'}
			\slsh{\eta'} \gamma_{d} \slsh{\eta'} \right. \nonumber \\
		& &  \left. \spop{'} \schm{'}{'}{'} \cschm{'}{'}{} \right\} .
\end{eqnarray}
Using the identities $\slsh{\eta} \gamma_{a} \slsh{\eta} = 2 \eta_{a}
\slsh{\eta} - a^{2} \gamma_{a}$
and $ \eta_{a} \Q{ab}{}{} = 0$ the above simplifies to
\begin{eqnarray}
	A_{0} & = & \FactorTwo \sum_{lm} \sum_{l'm'} G(l) G(l') \nonumber \\
		& & TR \left\{ \gamma^{c} \ids
			\cschm{'}{'}{} \Q{a}{\;\; c}{} \spop{} \schm{}{}{} \right. \nonumber \\
		& & \left.\; \; \; \gamma_{d} \idsx{'}
			 \cschm{}{}{'} \Q{d}{\;\; a}{'} \spop{'} \schm{'}{'}{'} \right\} .
\end{eqnarray}
The matrix elements may be written down by simply replacing the $\eta$-space
operators by their
$l$-space counterparts with due care given to the orthonormality conditions of
the harmonics.  This means
that the sum over $m$ indices gives a trace and that the net change of $l$
eigenvalue for the trace is zero.
For the specific case above one gets
\begin{eqnarray}
	A_{0} & = & \FactorTwo \sum_{l} \sum_{l'} G(l) G(l') \nonumber \\
		& & \sum_{\delta,\delta'} tr \left[ TR \left\{ \gamma^{c}
\Ql{'}{\delta'}{a}{\;\; c}
			\spopl{} \right. \right. \nonumber \\
		& & \left. \left. \; \; \; \gamma_{d} \Ql{}{\delta}{d}{\;\; a} \spopl{'}
\right\} \right]
			\delta_{l,l'+\delta'} \delta_{l',l+\delta} ,
\end{eqnarray}
where $tr\{ \ldots \}$ refers to the $m$ trace.  The condition on $l$ values
implies that the sum over
$(\delta,\delta')$ is only over the pairs which sum to zero.  Performing the
trace over gamma matrices,
using the identities in Appendix(\ref{gammaid}), leads to
\begin{eqnarray}
	A_{0} & = & \FactorThree \sum_{l} \sum_{l'} G(l) G(l') \nonumber \\
		& & \inda \sum_{\delta,\delta'} \left[
			(\Aterm)^{2} tr \left\{ \Ql{}{\delta}{d}{\;\; a}
			\Ql{'}{\delta'}{\;\; a}{d} \right\} \right. \nonumber \\
		& & \indb \mbox{} -(\Aterm) tr \left\{ \Ql{}{\delta}{d}{\;\; a}
\Ql{'}{\delta'}{ca}{}
			\tl{}{}{cd} \right\} \nonumber \\
		& & \indb \mbox{} -(\Aterm) tr \left\{ \Ql{}{\delta}{d}{\;\; a} \tl{'}{}{dc}
			\Ql{'}{\delta'}{ca}{} \right\} \nonumber \\
		& & \indb \mbox{} + 1 \inda tr \left\{ \Ql{}{\delta}{d}{\;\; a}
\tl{'}{e}{\;\; c}
			\Ql{'}{\delta'}{ca}{} \tl{}{}{ed} \right. \nonumber \\
			& & \indc \mbox{} + \Ql{}{\delta}{d}{\;\; a} \tl{'}{}{de}
				\Ql{'}{\delta'}{ca}{} \tl{}{\;\; e}{c} \nonumber \\
			& & \indc \left. \left. \mbox{} -\half \Ql{}{\delta}{d}{\;\; a} \tl{'}{}{ef}
				\Ql{'}{\delta'}{\;\; a}{d} \tl{}{ef}{} \right\} \right]
				\delta_{l,l'+\delta'} \delta_{l',l+\delta} ,
\end{eqnarray}
which using the identities in Appendix(\ref{OpId}) may be written in the
factorized form
\begin{eqnarray}
	A_{0} & = & \FactorThree \sum_{l} \sum_{l'} G(l) G(l') \nonumber \\
		& & \inda \sum_{\delta,\delta'} tr \left\{ \Ql{}{\delta}{d}{\;\; a}
			\spoplab{'}{a}{\;\; c} \right. \nonumber \\
		& & \indd . \Ql{'}{\delta'}{c}{\;\; e} \spoplab{}{e}{\;\; d} \nonumber \\
		& & \left. \indc \mbox{} - (n-2) \htlsq{} \delta_{\delta,0}
\delta_{\delta',0}
			\right\} \delta_{l,l'+\delta'} \delta_{l',l+\delta} .
\end{eqnarray}
Contracting the indices for each $(\delta,\delta')$ using the definitions
Eq.(\ref{Qlplus}),
Eq.(\ref{Qlminus}) and Eq.(\ref{Qlzeropl}) gives
\begin{eqnarray}
	(0,0) \inda & = & \inda (\nontwo - 2)^{2}
			tr \left\{ \vpl{}{0}{d}{\;\; a} \vpl{}{0}{a}{\;\; d} \right\} \nonumber \\
		& & \mbox{} + (l+\nontwo - 1)^{2} \frac{(l+n)^{2}}{(2l+n+1)^{2}}
			tr \left\{ \vpl{}{+1}{d}{\;\; a} \vpl{}{+1}{a}{\;\; d} \right\} \nonumber \\
		& & \mbox{} + (l+\nontwo)^{2} \frac{(l-1)^{2}}{(2l+n-3)^{2}}
                        tr \left\{ \vpl{}{-1}{d}{\;\; a} \vpl{}{-1}{a}{\;\; d}
\right\} \nonumber \\
		& & \mbox{} + (n-2)\, l(l+n-1) \, tr\{1\} \nonumber \\
		& = & \inda (\nontwo - 2)^{2} dim(l,1) \nonumber \\
		& & \mbox{} + (l+\nontwo - 1)^{2} \frac{(l+n)^{2}}{(2l+n+1)^{2}} dim(l+1,0)
\nonumber \\
		& & \mbox{} + (l+\nontwo)^{2} \frac{(l-1)^{2}}{(2l+n-3)^{2}} dim(l-1,0)
\nonumber \\
		& & \mbox{} + (n-2)\, l(l+n-1) dim(l,0) ,
\end{eqnarray}
\begin{eqnarray}
	(+2,-2) \inda & = & \inda -(l+\nontwo + 2)(l+\nontwo - 1)
			tr \left\{ \Ql{}{+2}{d}{\;\; a} \Ql{+2}{-2}{a}{\;\; d} \right\} \nonumber \\
		& = & \inda -(l+\nontwo + 2)(l+\nontwo - 1)
			\frac{(l+1)(l+n)}{(2l+n+1)^{2}} tr \left\{ \vpl{}{+1}{d}{\;\; d}
			\right\} \nonumber \\
		& = & \inda -(l+\nontwo + 2)(l+\nontwo - 1)
                        \frac{(l+1)(l+n)}{(2l+n+1)^{2}} dim(l+1,0) ,
\end{eqnarray}
\begin{eqnarray}
	(-2,+2) \inda & = & \inda -(l+\nontwo - 3)(l+\nontwo )
                        tr \left\{ \Ql{}{-2}{d}{\;\; a} \Ql{-2}{+2}{a}{\;\; d}
\right\} \nonumber \\
                & = & \inda -(l+\nontwo - 3)(l+\nontwo )
                        \frac{(l-1)(l+n-2)}{(2l+n-3)^{2}} tr \left\{
\vpl{}{-1}{d}{\;\; d}
                        \right\} \nonumber \\
                & = & \inda -(l+\nontwo - 3)(l+\nontwo )
                        \frac{(l-1)(l+n-2)}{(2l+n-3)^{2}} dim(l-1,0) .
\end{eqnarray}
and including the $l$ constraints, one gets
\begin{eqnarray}
	A_{0} & = & \FactorThree \sum_{l} \nonumber \\
		& & \inda \left\{ G(l)^{2}\, \left[ (\nontwo - 2)^{2} dim(l,1) \right.
\right. \nonumber \\
		& & \indc \mbox{} + (l+\nontwo - 1)^{2}
			\frac{(l+n)^{2}}{(2l+n+1)^{2}} dim(l+1,0) \nonumber \\
		& & \indc \mbox{} + (l+\nontwo)^{2}
			\frac{(l-1)^{2}}{(2l+n-3)^{2}} dim(l-1,0) \nonumber \\
		& & \left. \indc \mbox{} + (n-2)\, l(l+n-1) dim(l,0) \right] \nonumber \\
		& & \inda \mbox{} - G(l)\, G(l+2)\, (l+\nontwo + 2)(l+\nontwo - 1)
			\frac{(l+1)(l+n)}{(2l+n+1)^{2}} dim(l+1,0) \nonumber \\
		& & \left. \inda \mbox{} - G(l)\, G(l-2)\, (l+\nontwo - 3)(l+\nontwo )
			\frac{(l-1)(l+n-2)}{(2l+n-3)^{2}} dim(l-1,0) \right\} .
\end{eqnarray}

There are various ways\cite{ds2,ds3} to evaluate the $l$-sums obtained in this
way.  The method used here
is described in Appendix(\ref{lsums}) with the $dim(l,0)$ term above used as an
example.
Briefly it involves expanding
the terms of the sum in powers of $l$ and extracting the divergence as the pole
of the Riemann
$\zeta$-function.  Shifting the sum where necessary, the expansions
\begin{equation}
	\frac{\Gamma(l-\eps)}{\Gamma(l)} = l^{-\eps}\{ 1 + O(\eps) \}
\end{equation}
and
\begin{equation}
	(l+A)^{-N} = l^{-N} \left( 1 - {\case{N}/{Al}} + \cdots \right)
\end{equation}
allow the sum to be expressed in the form
\begin{eqnarray}
	X(\eps) & = & \sum_{l} \sum_{k} a_{k}(\eps) l^{-k-\eps} \nonumber \\
		& = & \sum_{k} a_{k}(\eps) \zeta(k+\eps) .
\end{eqnarray}
Taking the limit $\eps \rightarrow 0$ one obtains the pole term and a finite
part if required.
Since we are only concerned with the divergent part in what follows we have
\begin{eqnarray}
	X_{div} & = & a_{1}(\eps) \zeta(1+\eps) |_{\eps \approx 0} \nonumber \\
		& = & a_{1}(0) \left( \oneps \right) .
\end{eqnarray}

Dropping the terms which are proportional to $(n-4)$ we have for $A_{0}$,
\begin{eqnarray}
	A_{0} & = & \FactorThree \sum_{l} \nonumber \\
		& & \left\{ \left[ \frac{(l+n)}{(l+\nontwo)^{2}} -
\frac{(l+1)}{(l+\nontwo)(l+\nontwo+1)}
			\right] \frac{(l+n)}{(2l+n+1)^{2}} dim(l+1,0) \right. \nonumber \\
		& & \mbox{} + \left[ \frac{(l-1)}{(l+\nontwo-1)^{2}} -
\frac{(l+n-2)}{(l+\nontwo-1)(l+\nontwo-2)}
                        \right] \frac{(l-1)}{(2l+n-3)^{2}} dim(l-1,0) \nonumber
\\
		& & \left. \mbox{} + (n-2) \frac{l(l+n-1)}{(l+\nontwo)^{2}(l+\nontwo-1)^{2}}
			dim(l,0) \right\} \nonumber \\
		& = & \FactorThree \sum_{l} \nonumber \\
		& & \left\{ \frac{n}{2} \, \frac{(l+n)}{(l+\nontwo)^{2}(l+\nontwo+1)}
			\inda \frac{\Gamma(l+n)}{\Gamma(n) \Gamma(l+2)} \right. \nonumber \\
		& & \mbox{} - \frac{n}{2} \, \frac{(l-1)}{(l+\nontwo-1)^{2}(l+\nontwo-2)}
                        \inda \frac{\Gamma(l+n-2)}{\Gamma(n) \Gamma(l)}
\nonumber \\
		& & \left. \mbox{} + (n-2) \,
\frac{l(l+n-1)}{(l+\nontwo)^{2}(l+\nontwo-1)^{2}}
			\inda \dlz \right\} ,
\end{eqnarray}
which after a little work gives
\begin{eqnarray}
	A_{0} & = & \FactorFour \sum_{l} \left\{ 2(l+2) - 2(l-2) + 2 \frac{1}{(l+1)}
(l-1)(l+2)(2l+1) \right\}
			l^{-1-\eps} \nonumber \\
		& = & \FactorFour \left\{ 4 + 4 + 2(-3-3+2)\right\} \oneps = 0 .
\end{eqnarray}
This result is gratifying as the Ward Identity Eq.(\ref{PiWardIds}) says that
$A_{0}$ must be zero.

In order to calculate $A_{1}$ , note that since $\xonyasq{1}{2}(\eta -
\eta')^{2} =
1 - \xonyasq{\eta \cdot \eta'}{}$ and $A_{0}=0$ we can write from
Eq.(\ref{Pidist4}),
\begin{eqnarray}
	A_{1} & = & - {\case{1}/{n(n-1) \Omega_{n+1} a^{2}}} \idssp
			\left(\xonyasq{\eta \cdot \eta'}{} \right)
			\Pi^{a}_{\,\, a}(\eta,\eta') \nonumber \\
		& = & - \FactorFive \onasq \sum_{lm} \sum_{l'm'} G(l) G(l') \nonumber \\
                & & TR \left\{ \gamma^{c} \ids
                        \cschm{'}{'}{} \Q{a}{\;\; c}{} \eta^{b} \spop{}
\schm{}{}{} \right. \nonumber \\
                & & \left. \indb \gamma_{d} \idsx{'}
                        \cschm{}{}{'} \eta'_{b} \Q{d}{\;\; a}{'}
			\spop{'} \schm{'}{'}{'} \right\} \nonumber \\
		& = & - \FactorFive \onasq \sum_{l} \sum_{l'} G(l) G(l') \nonumber \\
		& & \sum_{\delta,\delta'} \sum_{\sigma, \sigma'}
			tr \left[ TR \left\{ \gamma^{c} \Ql{'}{\delta'}{a}{\;\; c}
			\etal{'+\delta'}{\sigma'}{b}{} \spopl{} \right. \right. \nonumber \\
		& & \left. \left. \indc \gamma_{d} \etal{}{\sigma}{}{b}
\Ql{+\sigma}{\delta}{d}{\;\; a}
			\spopl{'} \right\} \right] \nonumber \\
		& & \inde \delta_{l,l'+\sigma'+\delta'} \delta_{l',l+\sigma+\delta} \nonumber
\\
		& = & - \FactorFive \onasq \sum_{l} \sum_{l'} G(l) G(l') \nonumber \\
		& & \inda \sum_{\delta,\delta'} \sum_{\sigma,\sigma'}
			tr \left\{ \Ql{+\sigma}{\delta}{d}{\;\; a}
			\spoplab{'}{a}{\;\; c} \Ql{'}{\delta'}{c}{\;\; e} \right. \nonumber \\
		& & \indd . \etal{'+\delta'}{\sigma'}{b}{} \spoplab{}{e}{\;\; d}
			\etal{}{\sigma}{b}{} \nonumber \\
		& & \left. \indc \mbox{} - \half (n-2) \tl{'}{}{cd} \etal{'}{\sigma'}{b}{}
			\tl{}{cd}{} \etal{}{\sigma}{}{b} \delta_{\delta,0} \delta_{\delta',0}
			\right\} \nonumber \\
		& & \inde \delta_{l,l'+\sigma'+\delta'} \delta_{l',l+\sigma+\delta} .
\end{eqnarray}
Refering to the indentities in Appendix(\ref{OpId}) , it can be seen that the
non-zero values for
$(\sigma,\sigma')$ are $(+1,-1)$ and $(-1,+1)$ . Contracting the indices for
the two cases
respectively we get
\begin{eqnarray}
	A_{1} & = & - \FactorFive \sum_{l} \sum_{l'} G(l) G(l') \nonumber \\
		& & \left[ \frac{1}{(2l+n+1)} \sum_{\delta,\delta'}
			tr \left\{ \Ql{+1}{\delta}{d}{\;\; a}
                        \spoplab{'}{a}{\;\; c} \right. \right. \nonumber \\
		& & \indc \Ql{'}{\delta'}{c}{\;\; e}
			\left[ (\Aterm)(l+1)\kdelta{e}{\;\; d} - (l)\tl{+1}{e}{\;\; d} \right]
			\nonumber \\
		& & \left. \indd \mbox{} - (n-2)(l) \htlsq{+1} \delta_{\delta,0}
\delta_{\delta',0}
                        \right\} \delta_{l',l+\delta+1} \nonumber \\
		& & \mbox{} + \frac{1}{(2l+n-3)} \sum_{\delta,\delta'}
                        tr \left\{ \Ql{-1}{\delta}{d}{\;\; a}
                        \spoplab{'}{a}{\;\; c} \right. \nonumber \\
                & & \indc \Ql{'}{\delta'}{c}{\;\; e}
                        \left[ (\Aterm)(l+n-2)\kdelta{e}{\;\; d} -
(l+n-1)\tl{-1}{e}{\;\; d} \right]
                        \nonumber \\
                & & \left. \left. \indd \mbox{} - (n-2)(l+n-1) \htlsq{-1}
			\delta_{\delta,0} \delta_{\delta',0} \right\} \delta_{l',l+\delta-1} \right]
{}.
\end{eqnarray}
The above may be evaluated in the same way as for $A_{0}$ , and after some work
the result
\begin{equation}
	A_{1} = \FactorSix + O(1)
\end{equation}
is obtained.  Therefore we have
\begin{equation}
	\Pi^{ab}(\eta,\eta') = \FactorSix \pop{ab}{}(\eta) \delta(\eta - \eta') + O(1)
\end{equation}
and so the addition of the counterterm
\begin{equation}
	\delta S_{E} = \ids \left[ \FactorSixA \right] \xonyasq{1}{2} A_{a} \pop{ab}{}
A_{b}
\end{equation}
is required to cancel the divergence.  This gives to this order
\begin{equation}
	Z_{3} = 1 - \FactorSixA
\end{equation}
for the photon wave function renormalization constant which is identical to
that for flat space, as
expected.

\subsection{Fermion Self Energy}

The fermion self-energy, (FIG.\ref{fermfig}) is given by
\begin{equation}
	\Sigma(\eta,\eta') = (ie)^{2} \mu^{4-n} \Q{ac}{}{} \gamma_{c} S(\eta,\eta')
				\Q{bd}{}{'} \gamma_{d} E_{ba}(\eta',\eta) .
\end{equation}
We will first evaluate $B_{0}$ using Eq.(\ref{Sigd1}) and Eq.(\ref{Sigd2}) ,
\begin{eqnarray}
	B_{0} & = & \FactorSevenA \idssp TR\left\{ \Sigma(\eta,\eta') \right\}
\nonumber \\
		& = & - \FactorSeven \onasq \sum_{lm} \sum_{l'm'} G(l) E(l') \nonumber \\
		& & TR \left\{ \ids \cschm{'}{'}{} \Q{ac}{}{} \gamma_{c} \slsh{\eta}
			\spop{} \schm{}{}{} \right. \nonumber \\
		& & \left. \idsx{'} \cschm{}{}{'} \Q{bd}{}{'} \gamma_{d}
			\vctop{'}{}{ba}{'} \schm{'}{'}{'} \right\} \nonumber \\
		& = & 0 ,
\end{eqnarray}
since there is no way to connect $l$-eigenvalues as $\delta l =$ odd for the
first integral and
$\delta l$ = even for the second. Note, we cannot use the property $TR\{ odd \:
\# \: \gamma 's \}=0$
here as the trace of 5 $\gamma$ matrices is non-zero.

For the next term we use Eq.(\ref{Sigd3})
\begin{eqnarray}
	B_{1} - \nontwo B_{2} & = & \FactorSevenA \idssp TR\left\{ \slsh{\eta}
					\Sigma(\eta,\eta') \right\} \nonumber \\
		& = & \FactorSeven \sum_{lm} \sum_{l'm'} G(l) E(l') \nonumber \\
		& & TR \left\{ \ids \cschm{'}{'}{} \Q{ac}{}{} \gamma_{c}
			\spop{} \schm{}{}{} \right. \nonumber \\
		& & \left. \idsx{'} \cschm{}{}{'} \Q{bd}{}{'} \gamma_{d}
                        \vctop{'}{}{ba}{'} \schm{'}{'}{'} \right\} ,
\end{eqnarray}
performing the spinor trace and going to $l$-space we have
\begin{eqnarray}
        B_{1} - \nontwo B_{2} & = & \FactorEight \sum_{l,l'} G(l) E(l')
\nonumber \\
                & & \sum_{\delta, \delta'} tr \left\{ \Ql{'}{\delta'}{ac}{}
			\spoplab{}{c}{\;\; d} \right. \nonumber \\
		& & \inda \left. \Ql{}{\delta}{db}{} \vctopl{'}{}{ba} \right\} .
\end{eqnarray}
The operator $\vpl{'}{0}{}{ba}$ projects out all but the $\vpl{}{0}{}{}$ parts
of the
$Q$ matrices and the eigenvalue of $[ (\Aterm) - T]$ is $(\nontwo -2)$ on this
subspace.
Therefore we see immediately that the $(1-\alpha)$ term is finite.
We now have
\begin{eqnarray}
	B_{1} - \nontwo B_{2} & = & \FactorEight \alpha \sum_{l,l'} G(l) E(l')
\nonumber \\
		& & \sum_{\delta, \delta'} tr \left\{ \Ql{'}{\delta'}{ac}{}
\spoplab{}{c}{\;\; d}
			\Ql{}{\delta}{d}{\;\; a} \right\} \nonumber \\
		& & \indd \delta_{l',l+\delta} \delta_{l,l'+\delta'} ,
\end{eqnarray}
which may be summed in the usual way giving
\begin{equation}
	B_{1} - 2B_{2} = {\case{e^{2}}/{8\pi^{2}}} \alpha \left( \twooneps \right) .
\end{equation}
By rearranging the propagator $S(\eta,\eta')$ using Eq.(\ref{splprop2}) and
Eq.(\ref{Sigd3}) one can show that to this order
\begin{equation}
	B_{1} + \nontwo B_{2} = - ( B_{1} - \nontwo B_{2} )
\end{equation}
and so
\begin{equation}
	B_{1} = 0
\end{equation}
and
\begin{equation}
	B_{2} = - \FactorNine .
\end{equation}
Therefore we have
\begin{equation}
	\Sigma(\eta,\eta') = - \FactorNine \left( \slsh{D} - \xonyasq{n}{2}
\slsh{\eta}
				\right) \delta(\eta - \eta')
\end{equation}
which is cancelled by the counterterm
\begin{equation}
	\delta S_{E} = - \ids \left[ \FactorNine \right] \bar{\psi}
			\left( \slsh{D} - \xonyasq{n}{2} \slsh{\eta} \right) \psi
\end{equation}
and so
\begin{equation}
	Z_{2} = 1 - \FactorNine .
\end{equation}

\subsection{Vertex Function}

The irreducible vertex function may also be expanded as a distribution, but we
will consider
only the divergent part
\begin{equation}
	\Gamma^{a}(\eta_{1},\eta_{2},\eta) = -ie(\mu)\, C_{0} \Q{a}{\;\; b}{}
\gamma_{b}
			\delta(\eta - \eta_{1}) \delta(\eta - \eta_{2}) + \cdots
\end{equation}
and so
\begin{equation}
	C_{0} = \left[ -ie(\mu)\, 2^{n/2} n \vol \right]^{-1} \idsthree{_{1}}{_{2}}{}
		TR \left\{ \Gamma^{a}(\eta_{1},\eta_{2},\eta) \gamma_{a} \right\} .
\end{equation}
To this order we have (FIG.\ref{vertfig})
\begin{eqnarray}
	\Gamma^{a}(\eta_{1},\eta_{2},\eta) & = & (-1)[ie(\mu)]^{3} \Q{}{bd}{_{1}}
\gamma^{d}
		S(\eta_{1},\eta) \Q{a}{\;\; e}{} \gamma^{e} \nonumber \\
	& & \inda S(\eta,\eta_{2}) \Q{}{cf}{_{2}} \gamma^{f} E^{cb}(\eta_{2},\eta_{1})
,
\end{eqnarray}
and therefore
\begin{eqnarray}
	C_{0} & = & - \FactorTen \sum_{lm} \sum_{l'm'} \sum_{l'',m''} G(l)\, G(l')\,
E(l'') \nonumber \\
		& & TR \left\{
			\idsx{_{1}} \cschm{''}{''}{_{1}} \Q{}{bd}{_{1}} \gamma^{d} \spop{_{1}}
			\schm{}{}{_{1}} \right. \nonumber \\
		& & \inda
			\ids \cschm{}{}{} \Q{a}{\;\; e}{} \gamma^{e} \spop{}
			\schm{'}{'}{} \nonumber \\
		& & \left. \inda
			\idsx{_{2}} \cschm{'}{'}{_{2}} \Q{}{cf}{_{2}} \vctop{''}{cb}{}{_{2}}
			\schm{''}{''}{_{2}} \right\}
\end{eqnarray}
where we have again anticommuted the $\slsh{\eta}$ factors from the spinor
propagators through
the $\gamma^{e}$ matrix discarding the term proportional to $\eta^{e}$ .  Going
to
$l$-space and evaluating the trace we get
\begin{eqnarray}
	C_{0} & = & - \FactorEleven \sum_{l,l',l''} G(l)\, G(l')\, E(l'') \nonumber \\
		& & \sum_{\delta,\delta',\delta''} tr \left\{
			\Ql{''}{\delta''}{}{bd} \spoplab{}{d}{\;\; e} \right. \nonumber \\
		& & \inda
			\left[ 2\Ql{}{\delta}{e}{\;\; g} - n \kdelta{e}{\;\; g} \delta_{\delta,0}
\right]
			\spoplab{'}{g}{\;\; f} \nonumber \\
		& & \left. \inda
			\Ql{'}{\delta'}{f}{\;\; c} \vctopl{''}{cb}{} \right\} \nonumber \\
		& & \indb
			\delta_{l,l''+\delta''} \delta_{l',l+\delta} \delta_{l'',l'+\delta'}
\end{eqnarray}
where terms of $O(n-4)$ have been neglected.  Note again that the $(1-\alpha)$
term is also
finite due to the $\vpl{''}{0}{cb}{}$ operator .

This sum may be evaluated to give
\begin{equation}
	C_{0} = \FactorNine ,
\end{equation}
and so
\begin{equation}
	\Gamma^{a}(\eta_{1},\eta_{2},\eta) = -ie(\mu)\, \left[ \FactorNine
			\right] \Q{a}{\;\; b}{} \gamma_{b}
			\delta(\eta - \eta_{1}) \delta(\eta - \eta_{2}) + \cdots .
\end{equation}
This divergence is removed by the counterterm
\begin{equation}
	\delta S_{E} = \ids +ie(\mu) \left[ \FactorNine \right] \bar{\psi} \Q{a}{\;\;
b}{}
			\gamma_{b} \psi A_{a} ,
\end{equation}
and hence
\begin{equation}
	Z_{1} = 1 - \FactorNine .
\end{equation}
This gives the result $Z_{1}=Z_{2}$ which is required by the Ward Identity
Eq.(\ref{VWItwo}) .

\section{Discussion}

In this paper, we have constructed a consistent formulation of field theory on
$S_{4}$ and described
a method from which various one-loop amplitudes may be calculated.
The above results were presented as an illustration of how calculations may be
carried out
using this method.  The standard results of QED are derived, providing a
valuable check on the
validity of this approach, as well as allowing a direct comparision between
this and previous
work done in the area.  While much of the basic formalism has been
derived previously\cite{drum1,ds1,ds2,ds3} , we believe that the matrix element
approach offers a new way
to deal with some of the more troublesome aspects of previous calculations.  In
particular one
can easily handle the transverse part of the photon propagator which had made
the $(1-\alpha)$ gauge parts
difficult to calculate. However the main advantage of this method is the
ability to compute
functions which involve the contraction of indices across different $\eta$
integrals.
This tends to happen when one has derivative couplings such as those in scalar
electrodynamics.
To be more concrete, consider the amplitude
\begin{equation}
	X = \idssp D_{a}G(\eta,\eta')\, D'^{a}G(\eta',\eta)
\end{equation}
which can arise in a theory with scalar loops\cite{ds3}.  The brute force
approach is to write down the
propagators in configuration space, perform the derivatives and try do the
integrals obtained using
Geigenbauer polynomials and the like.  While this approach will work here, it
is cumbersome and relies
heavily on the ability to write down the configuration space form of the
propagators.  When we come
to a non-Abelian theory, we encounter for example, the gluon three-point
function.  This function involves
derivative couplings which act on vector propagators which hook onto the
vertex. Any calculation
involving this vertex, such as the vacuum polarization, will be difficult to
calculate even in the
Feynman gauge using configuration space methods, whereas if one follows the
matrix element prescription
the result can be obtained for a general gauge in a lengthy, but
straightforward way.

This calculational scheme provides all the tools necessary to evaluate matrix
elements of operators between two spherical harmonics.  For one-loop diagrams
this
is all that is required. However in a two-loop amplitude such as an interacting
vacuum
diagram, one may have three or more harmonics at the same point.  In this case
one
must perform a Clebsch-Gordon decomposition of the product in order to reduce
the integral
to one over only two harmonics.  This is technically difficult in general, but
for a special
class of propagator functions of the form $\left| \eta - \eta' \right|^{m}$ it
is possible
and the method will be described in a forthcoming paper.

It may be useful to incorporate
chiral fermions into the theory by appropriate inclusion of the $\slsh{\eta}$
operator, bearing in mind that one will need to be careful with the dimensional
regularization
when the trace of five $\gamma$ matrices in encountered.

The main application of this work is that is allows one to take any
sensible model one wishes and work out all the relevant renormalisation group
equations
connecting the coupling constants to the parameter $d$ in the effective action
Eq.(\ref{effact}).
By considering the constraints imposed by the wormhole hypothesis, these
equations may provide
information on the observed values of the constants of nature.

\appendix{Gamma Matrix Identities in $SO(n+1)$ }
\label{gammaid}
\begin{eqnarray}
	\left\{ \gamma_{a} , \gamma_{b} \right\} & = & 2 \kdelta{}{ab} \\
	\gamma_{a}^{\dagger} & = & \gamma_{a} \\
	\Sigma_{ab} & = & {\case{1}/{4}} \left[ \gamma_{a} , \gamma_{b} \right] \\
	\gamma_{a} \gamma^{a} & = & (n+1) I \\
	TR \left\{ I \right\} & = & 2^{n/2} \\
	TR \left\{ odd \; \# \; < 5 \right\} & = & 0 \\
	TR \left\{ \gamma_{a} \gamma_{b} \gamma_{c} \gamma_{d} \gamma_{e} \right\} & =
&
		4 \epsilon_{abcde} \;\; for \; n=4 \\
	TR \left\{ \gamma_{a} \gamma_{b} \right\} & = & 2^{n/2} \kdelta{}{ab} \\
	TR \left\{ \gamma_{a} \gamma_{b} \gamma_{c} \gamma_{d} \right\} & = &
		2^{n/2} \left( \kdelta{}{ab} \kdelta{}{cd} - \kdelta{}{ac} \kdelta{}{bd}
			+ \kdelta{}{ad} \kdelta{}{bc} \right)
\end{eqnarray}

\appendix{Operator and Matrix Element Identities}
\label{OpId}

\subsection{Commutators}

\begin{eqnarray}
	\left[ \hlsq , \lop{}{ab} \right]  & = &  0 \\
	\left[ \hlsq , \eta_{a} \right]  & = &  -n \eta_{a} + 2a^{2} D_{a} \\
	\left[ \hlsq , D_{a} \right]  & = &  - \xonyasq{2}{} \eta_{a}(\hlsq) +
(n-2)D_{a} \\
	\left[ \lop{}{ab} , \lop{}{cd} \right]  & = &  \kdelta{}{cb} \lop{}{ad} -
\kdelta{}{ca} \lop{}{bd}
				+ \kdelta{}{bd} \lop{}{ca} - \kdelta{}{ad} \lop{}{cb} \\
	\left[ \lop{}{ab} , \Qop{}{cd} \right]  & = &  \kdelta{}{cb} \Qop{}{ad} -
\kdelta{}{ca} \Qop{}{bd}
                                + \kdelta{}{bd} \Qop{}{ca} - \kdelta{}{ad}
\Qop{}{cb} \\
	\left[ \lop{}{ab} , \vpop{}{\lambda}{}{cd} \right]  & = &  \kdelta{}{cb}
				 \vpop{}{\lambda}{}{ad} - \kdelta{}{ca} \vpop{}{\lambda}{}{bd}
                                + \kdelta{}{bd} \vpop{}{\lambda}{}{ca} -
\kdelta{}{ad} \vpop{}{\lambda}{}{cb} \\
	\left[ \lop{}{ab} , \eta_{c} \right]  & = &  \eta_{a} \kdelta{}{bc} - \eta_{b}
\kdelta{}{ac} \\
	\left[ \lop{}{ab} , D_{c} \right]  & = &  D_{a} \kdelta{}{bc} - D_{b}
\kdelta{}{ac} \\
	\left[ D_{a} , D_{b} \right]  & = &  \onasq \lop{}{ab} \\
	\left[ D_{a} , \eta_{b} \right]  & = &  \Qop{}{ab} \\
	\left[ D_{a} , \Qop{}{bc} \right]  & = &  - \onasq \left( \Qop{}{ab} \eta_{c}
					+ \Qop{}{ac} \eta_{b} \right)
\end{eqnarray}

\subsection{Miscellaneous Identites}

\begin{eqnarray}
	\eta_{a} D_{b} - \eta_{b} D_{a}  & = &  \lop{}{ab} \\
	\pop{}{ab}  & = &  (\hlsq) \kdelta{}{ab} + \lop{}{ac} \lop{c}{\;\; b} - (n-1)
\lop{}{ab} \\
	F_{ab}  & = &  \lop{}{ac} \lop{c}{\;\; b} - (n-1) \lop{}{ab} + (n-2)
\kdelta{}{ab} \\
	\eta_{a} \lop{a}{\;\; b}  & = &  a^{2} D^{b} \\
	\lop{a}{\;\; b} \eta^{b}  & = &  n \eta^{a} - a^{2} D^{a} \\
	D_{a} \lop{a}{\;\; b}  & = &  -\onasq \eta_{b} (\hlsq) + (n-1) D_{b} \\
	\eta_{a} \lop{}{bc} + \eta_{b} \lop{}{ca} + \eta_{c} \lop{}{ab}  & = &  0 \\
	D_{a} \lop{}{bc} + D_{b} \lop{}{ca} + D_{c} \lop{}{ab}  & = &  0 \\
	\lop{a}{\;\; b} D^{b}  & = &  +\onasq \eta^{a} (\hlsq) + D^{a} \\
	\Qop{}{ab} \lop{b}{\;\; c} - \Qop{}{cb} \lop{b}{\;\; a}  & = &  \lop{}{ac} \\
	\lop{}{ab} \Qop{b}{\;\; c} - \lop{}{cb} \Qop{b}{\;\; a}  & = &  \lop{}{ac} \\
	\lop{ab}{} \Qop{\;\; c}{b} \lop{\;\; d}{c} - \lop{db}{} \Qop{\;\; c}{b}
\lop{\;\; a}{c}  & = &  0 \\
	\lop{ab}{} \Qop{\;\; c}{b} \lop{}{ca}  & = &  - \hlsq \\
	\Qop{ab}{} \lop{}{bc} \Qop{}{ab}  & = &  n \lop{}{bc}
\end{eqnarray}

\subsection{Matrix Identities}

The basic building blocks are the $U$ and $V$ matrices from which all others
may be derived.
\begin{eqnarray}
	\tl{}{}{ab} \vl{}{b}{} & = & -l \; \vl{}{}{a} \\
	\tl{}{}{ab} \ul{}{b}{} & = & (l+n-1) \; \ul{}{}{a} \\
	\ul{+1}{a}{} \tl{}{}{ab} & = & -l \;\ul{+1}{}{b} \\
	\vl{-1}{a}{} \tl{}{}{ab} & = & (l+n-1) \; \vl{-1}{}{b} \\
	\vl{}{}{a} \ul{+1}{a}{} & = & -\onasq (l+n-1)(2l+n+1) \\
	\ul{}{}{a} \vl{-1}{a}{} & = & -\onasq \, l \, (2l+n-3) \\
	\vl{}{}{a} \vl{+1}{a}{} & = & 0 \\
	\ul{}{}{a} \ul{-1}{a}{} & = & 0 \\
	\vl{}{}{a} \vl{+1}{}{b} - \vl{}{}{b} \vl{+1}{}{a} & = & 0 \\
	\ul{}{}{a} \ul{-1}{}{b} - \ul{}{}{b} \ul{-1}{}{a} & = & 0 \\
	\vl{}{}{a} \ul{+1}{}{b} - \vl{}{}{b} \ul{+1}{}{a} & = & \onasq (2l+n+1)\,
\tl{}{}{ab} \\
        \ul{}{}{a} \vl{-1}{}{b} - \ul{}{}{b} \vl{-1}{}{a} & = & -\onasq
(2l+n-3)\, \tl{}{}{ab} \\
	\tl{}{}{ab} \vl{}{}{c} - \vl{}{}{c} \tl{+1}{}{ab} & = &
			\vl{}{}{a} \kdelta{}{bc} - \vl{}{}{b} \kdelta{}{ac} \\
        \tl{}{}{ab} \ul{}{}{c} - \ul{}{}{c} \tl{-1}{}{ab} & = &
                        \ul{}{}{a} \kdelta{}{bc} - \ul{}{}{b} \kdelta{}{ac} \\
	\tl{}{}{ab} \vl{}{}{c} + \tl{}{}{bc} \vl{}{}{a} + \tl{}{}{ca} \vl{}{}{b} & = &
0 \\
	\tl{}{}{ab} \ul{}{}{c} + \tl{}{}{bc} \ul{}{}{a} + \tl{}{}{ca} \ul{}{}{b} & = &
0 \\
\end{eqnarray}

The $Q$ identities follow from their operator counterparts.
\begin{eqnarray}
	\tl{}{}{ab} \Ql{}{\delta}{}{cd} - \Ql{}{\delta}{}{cd} \tl{+\delta}{}{ab}
	& = & \kdelta{}{cb} \Ql{}{\delta}{}{ad} - \kdelta{}{ca} \Ql{}{\delta}{}{bd}
\nonumber \\
	& & \mbox{} + \kdelta{}{bd} \Ql{}{\delta}{}{ca} - \kdelta{}{ad}
\Ql{}{\delta}{}{cb}
\end{eqnarray}
\begin{eqnarray}
	\Ql{}{\delta}{\;\; b}{a} \tl{+\delta}{}{bc} -\Ql{}{\delta}{\;\; b}{c}
\tl{+\delta}{}{ba}
	& = & \tl{}{}{ac} \delta_{\delta,0} \\
	\tl{}{\;\; b}{a} \Ql{}{\delta}{}{bc} -\tl{}{\;\; b}{c} \Ql{}{\delta}{}{ba}
	& = & \tl{}{}{ac} \delta_{\delta,0} \\
	\tl{}{\;\; b}{a} \Ql{}{\delta}{\;\; c}{b} \tl{+\delta}{}{cd}
		- \tl{}{\;\; b}{d} \Ql{}{\delta}{\;\; c}{b} \tl{+\delta}{}{ca}
	& = & 0  \\
	\tl{}{\;\; b}{a} \Ql{}{\delta}{\;\; c}{b} \tl{+\delta}{\;\; a}{c} & = & -
\htlsq{} \; \delta_{\delta,0} \\
	\Ql{}{\delta}{}{ab} \tl{+\delta}{cd}{} \Ql{+\delta}{\delta'}{ab}{} & = &
		n \, \tl{}{cd}{} \delta_{\delta,0} \; \delta_{\delta',0}
\end{eqnarray}
Expressing $\Ql{}{\delta}{}{ab}$ in terms of projection operators and the $U$
and $V$ matrices
gives the eigenvalue relations
\begin{eqnarray}
	\tl{}{}{ab} \Ql{}{+2}{b}{\;\; c} & = & -l \Ql{}{+2}{}{ac} \\
	\tl{}{}{ab} \Ql{}{0}{b}{\;\; c} & = & \vpl{}{0}{}{ac}
		- {\case{l(l+n)}/{(2l+n+1)}} \vpl{}{+1}{}{ac}
		+ {\case{(l+n-1)(l-1)}/{(2l+n-3)}} \vpl{}{-1}{}{ac} \\
	& = & \Ql{}{0}{}{ab} \tl{}{b}{\;\; c} \\
	\tl{}{}{ab} \Ql{}{-2}{b}{\;\; c} & = & (l+n-1) \Ql{}{-2}{}{ac} \\
	\Ql{}{+2}{}{ab} \tl{+2}{b}{\;\; c} & = & (l+n+1) \Ql{}{+2}{}{ac} \\
	\Ql{}{-2}{}{ab} \tl{-2}{b}{\;\; c} & = & -(l-2) \Ql{}{-2}{}{ac} \\
\end{eqnarray}
and the product relations
\begin{eqnarray}
	\Ql{}{+2}{}{ab} \Ql{+2}{-2}{b}{\;\; c} & = &
{\case{(l+1)(l+n)}/{(2l+n+1)^{2}}}
		\vpl{}{+1}{}{ac} \\
	\Ql{}{-2}{}{ab} \Ql{-2}{+2}{b}{\;\; c} & = &
{\case{(l-1)(l+n-2)}/{(2l+n-3)^{2}}}
                \vpl{}{-1}{}{ac} \\
	\Ql{}{0}{}{ab} \Ql{}{0}{b}{\;\; c} & = & \vpl{}{0}{}{ac}
                + {\case{(l+n)^{2}}/{(2l+n+1)^{2}}} \vpl{}{+1}{}{ac}
                + {\case{(l-1)^{2}}/{(2l+n-3)^{2}}} \vpl{}{-1}{}{ac} \\
	\Ql{}{0}{}{ab} \Ql{}{+2}{b}{\;\; c} & = & {\case{(l+n)}/{(2l+n+1)}}
\Ql{}{+2}{}{ac} \\
	\Ql{}{0}{}{ab} \Ql{}{-2}{b}{\;\; c} & = & {\case{(l-1)}/{(2l+n-3)}}
\Ql{}{-2}{}{ac} \\
	\Ql{-2}{+2}{}{ab} \Ql{}{0}{b}{\;\; c} & = & {\case{(l-1)}/{(2l+n-3)}}
\Ql{-2}{+2}{}{ac} \\
	\Ql{+2}{-2}{}{ab} \Ql{}{0}{b}{\;\; c} & = & {\case{(l+n)}/{(2l+n+1)}}
\Ql{+2}{-2}{}{ac} \\
	\Ql{}{+2}{}{ab} \Ql{+2}{+2}{b}{\;\; c} & = & 0 \\
	\Ql{}{-2}{}{ab} \Ql{-2}{-2}{b}{\;\; c} & = & 0  .
\end{eqnarray}
Since the $\etal{}{\sigma}{}{a}$ and $\Dl{}{\sigma}{}{a}$ matrices are
proportional to
the $U$ and $V$ matrices similar identities hold.  Those of use here are
\begin{eqnarray}
	\etal{}{+1}{a}{} \etal{+1}{-1}{}{a} & = & a^{2} \, {\case{(l+n-1)}/{(2l+n-1)}}
\\
	\etal{}{-1}{a}{} \etal{-1}{+1}{}{a} & = & a^{2} \, {\case{(l)}/{(2l+n-1)}} \\
	\etal{}{+1}{a}{} \tl{+1}{}{bc} \etal{+1}{-1}{}{a} & = &
		a^{2} \, {\case{(l+n)}/{(2l+n-1)}} \tl{}{}{bc} \\
	\etal{}{-1}{a}{} \tl{-1}{}{bc} \etal{-1}{+1}{}{a} & = &
		a^{2} \, {\case{(l-1)}/{(2l+n-1)}} \tl{}{}{bc} \\
	\etal{}{+1}{a}{} \etal{+1}{+1}{}{a} & = & 0 \\
	\etal{}{-1}{a}{} \etal{-1}{-1}{}{a} & = & 0 \\
	\etal{}{+1}{a}{} \tl{+1}{}{bc} \etal{+1}{+1}{}{a} & = & 0 \\
	\etal{}{-1}{a}{} \tl{-1}{}{bc} \etal{-1}{-1}{}{a} & = & 0 .
\end{eqnarray}
Relevant traces are
\begin{eqnarray}
	tr \left\{ (1)_{l} \right\} & = & dim(l,0) \\
	tr \left\{ \vpl{}{+1}{a}{\;\; a} \right\} & = & dim(l+1,0) \\
	tr \left\{ \vpl{}{0}{a}{\;\; a} \right\} & = & dim(l,1) \\
	tr \left\{ \vpl{}{-1}{a}{\;\; a} \right\} & = & dim(l-1,0) .
\end{eqnarray}

\appendix{Evaluation of $l$-Sums}
\label{lsums}
Consider a sum of the form
\begin{eqnarray}
	A & = & \sum_{l} G(l)^{2} \, l(l+n-1) \, dim(l,0) \nonumber \\
	& = & \sum_{l} \frac{l(l+n-1)}{(l+\nontwo)^{2} (l+\nontwo -1)^{2}} \dlz
\end{eqnarray}
which arises in the Vacuum Polarization calculation.  In order to simplify the
calculation, it
is convienient to shift the sum so that the term in the denominator $(l+\nontwo
-1)$
becomes $(l+\nontwo -2) = l + O(\eps)$ eliminating the need to Taylor expand
this term.
Shifting the sum changes the result by a finite amount which we will ignore.
This is
useful in general to reduce the number of terms in the denominator which may
need to be
expanded.  Therefore we have for $A$
\begin{eqnarray}
	A & = & {\case{1}/{\Gamma(n)}} \sum_{l} \frac{(l-1)(l+n-2)}{(l+\nontwo -1)^{2}
(l+\nontwo -2)^{2}}
		(2l+n-3) \; \frac{\Gamma(l+n-2)}{\Gamma(l)} \nonumber \\
	& = & {\case{1}/{\Gamma(n)}} \sum_{l}
\frac{(l-1)(l+n-2)(l+n-3)(l+n-4)(2l+n-3)}
		{(l+\nontwo -1)^{2} (l+\nontwo -2)^{2}}
		\; \frac{\Gamma(l+n-4)}{\Gamma(l)} \nonumber \\
	& = & {\case{1}/{\Gamma(4-\eps)}} \sum_{l}
\frac{(l-1)(l+2-\eps)(l+1-\eps)(l-\eps)(2l+1-\eps)}
		{(l+1-\half \eps)^{2} (l-\half \eps)^{2}}
		\; \frac{\Gamma(l-\eps)}{\Gamma(l)} \nonumber \\
	& = & {\case{1}/{\Gamma(4)}} \sum_{l} \left[ \frac{1}{l(l+1)} (l-1)(l+2)(2l+1)
+ O(\eps) \right]
		l^{-\eps} \left[ 1 + O(\eps) \right] \nonumber \\
	& = & {\case{1}/{\Gamma(4)}} \sum_{l} \left[ 1 - {\case{1}/{l}} +
{\case{1}/{l^{2}}} \cdots \right]
		\left[ 2l^{3} + 3l^{2} - 3l - 2 \right] l^{-2-\eps} + O(\eps) \nonumber \\
	& = & {\case{1}/{6}} \left[ -3-3+2 \right] \oneps + \, finite \nonumber \\
	& = & -{\case{2}/{3}} \oneps + \, finite
\end{eqnarray}
using
\begin{equation}
	\frac{\Gamma(l-\eps)}{\Gamma(l)} = l^{-\eps} \left\{ 1 + \eps
		\left[ {\case{1}/{2l}} + {\case{1}/{12l^{2}}} - {\case{1}/{120l^{4}}} \cdots
		\right] \right\} ,
\end{equation}
keeping only terms of lowest order in $\eps$
and taking $\eps \rightarrow 0$ with $\zeta(1+\eps) \sim \oneps$.

\newpage
\figure{Vacuum Polarization. \label{vacfig}}
\figure{Fermion Self Energy. \label{fermfig}}
\figure{Vertex Function. \label{vertfig}}


\newpage
\begin{references}

\bibitem[1]{col1}S.Coleman, Nucl.\ Phys.\ {\bf B307}, 867(1988).
\bibitem[2]{col2}S.Coleman, Nucl.\ Phys.\ {\bf B310}, 643(1988).
\bibitem[3]{shore4}G.M.Shore, Phys.\ Rev.\ D{\bf 21}, 2226(1980).
\bibitem[4]{grin1}B.Grinstein and M.Wise, Phys.\ Lett.\ {\bf 212B}, 407(1988).
\bibitem[5]{grin2}B.Grinstein and C.T.Hill, Phys.\ Lett.\ {\bf 220B},
520(1989).
\bibitem[6]{drum1}I.T.Drummond, Nucl.\ Phys.\ {\bf B94}, 115(1975).
\bibitem[7]{ds1}I.T.Drummond and G.M.Shore, Ann.\ Phys.\ {\bf 117}, 89(1979).
\bibitem[8]{ds2}G.M.Shore, Ann.\ Phys.\ {\bf 117}, 121(1979).
\bibitem[9]{ds3}G.M.Shore, Ann.\ Phys.\ {\bf 128}, 376(1980).
\bibitem[10]{adler}S.L.Adler, Phys.\ Rev.\ D{\bf 6}, 3445(1972).
\bibitem[11]{itzub}C.Itzykson and J.B.Zuber, Quantum Field Theory,
(M$^{c}$Graw-Hill, 1985).
\bibitem[12]{drum2}I.T.Drummond, Phys.\ Rev.\ D{\bf 19}, 1123(1979).

\end{references}
\end{document}